\UseRawInputEncoding
\documentclass{article}

\usepackage{lineno,hyperref}
\modulolinenumbers[5]

\usepackage{times}
\usepackage{soul}
\usepackage{url}
\usepackage{hyperref}
\usepackage[small]{caption}
\usepackage{graphicx}
\usepackage{amsmath}
\usepackage{amsthm}
\usepackage{booktabs}
\usepackage{algorithm}
\usepackage{algorithmic}
\usepackage{amsfonts}
\usepackage{amsmath}
\usepackage{amssymb}
\usepackage{amsthm}

\usepackage{xspace}
\usepackage{xcolor}
\usepackage{graphicx}

\usepackage{multicol}
\usepackage{tikz}
\usetikzlibrary{arrows,automata}

\newtheorem{theorem}{Theorem}

\newtheorem{lemma}[theorem]{Lemma}

\newtheorem{definition}[theorem]{Definition}
\newtheorem{example}[theorem]{Example}

\newcommand{\Agt}{\mathsf{Agt}}
\newcommand{\Loc}{\mathsf{Loc}}
\newcommand{\loc}{\mathit{l}}
\newcommand{\init}{\loc_{\mathsf{init}}}
\newcommand{\Act}{\mathsf{Act}} 
\newcommand{\act}{\mathsf{act}} 
\newcommand{\Obs}{\mathsf{Obs}}
\newcommand{\Kn}[1]{\mathsf{Kn} (#1)}
\newcommand{\Inf}[1]{\mathsf{KS} (#1)}
\newcommand{\str}[1]{\mathsf{str} (#1)}
 \newcommand{\G}{\mathbf{G}}  

 \newcommand{\upd}[3]{\mathsf{update}(#1, #2, #3)}
 
\newcommand{\obj}{\Gamma} 
\newcommand{\fplay}{\pi}   
\newcommand{\play}{\pi}   
\newcommand{\hist}{\mathsf{h}} 
\newcommand{\last}{\mathsf{l}} 
\newcommand{\reach}{\mathsf{R}} 
\newcommand{\safe}{\mathsf{S}} 
\newcommand{\agone}{{\mathsf{a}_1}} 
\newcommand{\agtwo}{{\mathsf{a}_2}} 
\newcommand{\agi}{\mathsf{i}}

\newcommand{\agn}{{\mathsf{a}_n}} 
    \newcommand{\aga}{\mathbf{a}}
    \newcommand{\agb}{\mathbf{b}}

\newcommand{\strat}{\alpha}
\newcommand{\moore}{\mathsf{M}}

\newcommand{\magii}{\text{MAGIIAN}\xspace}

\newcommand{\defstyle}[1]{\textbf{#1}}

\newcommand{\comment}[1]{}
\newcommand{\defeq}{\stackrel{\mathsf{def}}{=}}
\newcommand{\defequiv}{\stackrel{\mathsf{def}}{\Longleftrightarrow}}

\newcommand{\set}[1]{\left\{ #1 \right\}}
\newcommand{\setdef}[2]{\left\{ #1 \>\middle|\> #2 \right\}}
\newcommand{\setdefn}[2]{\{ #1 \mid #2 \}}
\newcommand{\tuple}[1]{\left< #1 \right>}

\newcommand{\obsi}[2]{\mathsf{obs}_{#1} (#2)}
\newcommand{\obi}[2]{\mathsf{ob}_{#1} (#2)}

\title{Knowledge-Based Strategies for Multi-Agent 
       Teams Playing Against Nature} 

\author{
Dilian Gurov \\ KTH Royal Institute of Technology, Stockholm \\ dilian@kth.se  
\and
Valentin Goranko  \\ Stockholm University \\ valentin.goranko@philosophy.su.se
\and
Edvin Lundberg \\ Rocker AB \\ edvin\_lundberg@msn.com}

\begin{document}

\maketitle

\begin{abstract}
We study teams of agents that play against Nature towards achieving a common objective. The agents are assumed to have imperfect information due to partial observability, and have no communication during the play of the game. 
We propose a natural notion of \emph{higher-order knowledge} of agents. Based on this notion, we define a class of knowledge-based strategies, and consider the problem of synthesis of strategies of this class.
We introduce a multi-agent extension, MKBSC, of the well-known \emph{knowledge-based subset construction} applied to such games. Its iterative applications turn out to compute higher-order knowledge of the agents. We show how the MKBSC can be used for the design of knowledge-based strategy profiles, and investigate the transfer of existence of such strategies between the original game and in the iterated applications of the MKBSC, under some natural assumptions. 
We also relate and compare the ``intensional'' view on knowledge-based strategies based on explicit knowledge representation and update, with 
the ``extensional'' view on finite memory strategies based on finite transducers and show that, in a certain sense, these are equivalent.
\end{abstract}

\textbf{Keywords:}
multi-agent games, 
imperfect information, 
higher-order knowledge, 
knowledge-based strategies, 
strategy synthesis, 
Dec-POMDP


\section{Introduction}
\label{sec:introduction}

In this work we explore the strategy synthesis problem 
for teams (or coalitions) of agents that have to accomplish a given common 
objective, while acting under imperfect information and under various 
other natural assumptions. 
In particular, we are interested in the notion of knowledge of agents in 
that context and how it affects the strategic abilities of a team. 

When attempting to achieve an objective, intelligent agents act upon their knowledge: about the structure of the game itself, the history of the play so far, the other agents' strategies, observations, and actions. 
Knowledge has various aspects, but in the context of the present study the term refers to information that is structured and represented in a suitable way to be used by an agent for deciding on its actions towards achieving an objective. This knowledge can be ``static'', i.e., about the game structure, or ``dynamic'', i.e., about the play of the game.  While the static knowledge can be assumed as ``built'' into the agents' brain or design, the dynamic knowledge is re-computed and, if necessary, stored on-the-fly during the play.  

\paragraph{Motivation}
Strategy\footnote{Also called ``policy'' in the literature 
on planning and Dec-POMDP.}
 synthesis for teams of agents is a complex problem. In 
general, if no bound is put on the size of the memory of
the agents, the strategy synthesis problem is undecidable 
for coalitions of two or more agents in the presence of imperfect 
information, even for some basic classes of objectives
(see, e.g., \cite{DBLP:conf/focs/PnueliR90}).

When information is imperfect, agents typically need to
maintain and use a finite abstraction of the history in order
to be able to achieve the objective. We refer to this 
information, suitably structured for use, as ``(dynamic) knowledge''.
By ``knowledge states'' for an agent, in our context we mean sets of locations which the agent considers currently possible to be the actual location.  
We call strategies that are directly based on knowledge,
i.e., strategies that map knowledge states to actions
and update the knowledge state during play, ``knowledge-based 
strategies''. Such strategies can be attractive, since they 
are convenient for play, and are natural to explain to humans. 

To achieve certain objectives, agents may even have to 
maintain ``higher-order'' knowledge (i.e., knowledge about each 
other's knowledge). Intuitively, the higher the order (or nesting 
depth) of knowledge, the higher the strategic abilities of 
the coalition. 

For a bounded order of knowledge, the space of potential
knowledge states is finite and the synthesis problem of knowledge-based
strategies becomes decidable. It is this class of strategies
and their synthesis that we investigate here. 

\paragraph{Approach}
We study the above problem in the context of \emph{multi-agent 
games with imperfect information against Nature} (or \magii for short).
We make the following assumptions on the games and assume that 
they are common knowledge amongst all agents:
\begin{enumerate}
\itemsep0em
\item
the game arena is discrete, finite, and known to the agents,
\item
certain game states are indistinguishable for certain agents,
thus modelling the ``imperfect information'', 
\item
the agents cooperate, i.e., they are all in one team
playing against Nature,
\item
the agents may or may not see each others' actions,
\item
the agents cannot communicate with each other,
\item
the agents may or may not know each others' strategies. 
\end{enumerate}

We argue that the less the agents know or observe, i.e., the
higher their uncertainty is about the current state-of-affairs, the higher 
the impact is of maintaining higher-order knowledge. Thus, the case 
where agents cannot observe each others' actions and do not 
know each others' strategies is a natural starting point for 
studying also the less restricted cases, which we will discuss
below. 

As explained above, we only consider knowledge representations
with bounded memory (since the unbounded case gives rise to
undecidability results). However, within that class there is 
no a priori best choice of knowledge representation.
One may choose, for instance, to use the memory to remember the 
last~$n$ observed game locations; but most generally, the memory is 
used to compute and maintain some abstraction over the observed history
of locations. 

Inspired by a subset construction on single-agent games against
Nature, namely the Knowledge-Based Subset Construction
(or KBSC for short), which reduces games with imperfect information in 
a strategy-preserving fashion to ``expanded'' games of perfect information
(see, e.g., \cite{DBLP:journals/jcss/Reif84,DBLP:journals/lmcs/RaskinCDH07}),
we choose a representation based on sets of game locations.
The semantic interpretation of such a set is ``the best
estimate the agent can make about the current state-of-affairs''.
In the single-agent case this representation turns out to be 
sufficient for the class of parity objectives, as shown
in~\cite{DBLP:journals/lmcs/RaskinCDH07}.
Then, we represent higher-order knowledge by nesting 
recursively such sets of locations in a suitable fashion.

Also inspired by the KBSC, we investigate the correspondence
between knowledge-based strategies and memoryless, observation-based 
strategies in expanded games resulting from applying a generalised, 
multi-agent version of the KBSC, which we introduce here and call the 
MKBSC. The locations of the expanded games are conceptually joint
knowledge states of the agents. 
We call these two views on strategies the ``intensional'' and
the ``extensional'' view, respectively. The MKBSC can be iterated, 
essentially computing higher-order knowledge (i.e., incrementing 
the order of knowledge with each iteration). 

The correspondence between the two views is useful in several ways.
First, one can reduce the synthesis problem of knowledge-based 
strategies to the synthesis problem of memoryless,
observation-based strategies. Furthermore, by virtue of
the MKBSC construction, the individual observation-based 
memoryless strategies in the expanded games are simultaneously
memoryless strategies in single-agent games of perfect 
information that are intermediate games produced while 
computing the expansions. This can serve as the basis for
the design of efficient knowledge-based strategy synthesis 
algorithms, since algorithmic strategy synthesis for the latter class 
is well-studied (see e.g.~\cite{DBLP:books/cu/11/0001R11}).
Second, while strategy synthesis is more conveniently performed on 
the expanded games, once synthesised, the strategies can 
be presented to the agents as knowledge-based strategies, 
without the need for storing the expanded games, but by
recomputing the knowledge in the course of the play (i.e.,
on-the-fly). 
And third, there is a phenomenon that 
manifests itself much more explicitly in the extensional view: for
some games, the iterated MKBSC ``stabilises'', producing
isomorphic games from some iteration on. In the intensional
view, stabilisation corresponds to the existence of a
finite knowledge representation that contains the higher-order 
knowledge of the agents of \emph{any} order. Thus, for games on which 
the iterated construction eventually stabilises, this opens up 
the possibility to reason about common knowledge.

\paragraph{Contributions}
In this work we offer the following results and contributions.
First, 
we propose a formal notion of higher-order knowledge with a 
representation and semantic interpretation, and a notion of 
knowledge update.
Based on this, 
we provide a formal notion of knowledge-based strategies.
Next, we develop a 
generalisation of the KBSC to the case of multiple agents,
as a scheme that can be reused for other similar ``expansions''.
The construction in effect computes knowledge, and its
iteration computes higher-order knowledge. 
For this construction, we show
a strategy preservation result for perfect recall strategies 
with respect to reachability and safety objectives. The proof of this
result is constructive, and also reveals how to preserve 
finite-memory strategies. 
We then establish
a formal relationship between memoryless observation-based 
strategies in the expanded games, knowledge-based strategies 
in the original games, and the corresponding finite-memory 
strategies in the original games, and the equivalence of the 
latter two. With this we also exhibit formally the duality
between the intensional and the extensional views. 
From this correspondence, we obtain a reduction of the 
synthesis problem of knowledge-based strategies to that of 
memoryless observation-based strategies in expanded games. 
Then, we sketch
a heuristic for strategy synthesis, exploiting that the 
individual observation-based memoryless strategies in the 
expanded multi-agent games of imperfect information are 
simultaneously strategies in single-agent games of perfect 
information, which are intermediate games produced while 
computing the expansions. 
Further, we give a formal meaning to the statement that 
the higher the order of knowledge, the higher the strategic 
abilities of the team, and argue that this indeed is
the case here. 
(However, this increase is not strict, as will  be explained further.)
Finally, we establish that for some games the iterated MKBSC stabilises, 
in the sense that from some iteration on it results in isomorphic 
games. One implication of this is that
for stabilising games, the problem of existence of a winning 
knowledge-based strategy without a given bound on the knowledge 
nesting depth, is decidable. 

\paragraph{Related work}
The present work is in the intersection of several major research areas, incl.,  
decentralised cooperative decision making (often modelled by decentralised POMDPs), multi-agent planning, knowledge-based programs, games with imperfect information and strategy synthesis in them, etc. 
There is a huge body of more or less related literature, which we cannot possibly survey in any reasonable degree of detail here. So we only mention and briefly discuss some of the conceptually and technically closest works to ours and provide extensive, yet inevitably incomplete, lists of relevant references for these in Section~\ref{sec:related-work}. 

\paragraph{Structure}

The paper is organised as follows.
Section~\ref{sec:Synthesis} presents the strategy synthesis problem 
studied here. In particular, we show a motivating example where 
one needs knowledge of at least second-order in order to 
achieve the given objective. 
In Section~\ref{sec:MAGII} we define formally \magii, the formal
object of our study, and recall some standard notions from
the theory of games over finite graphs. 
Section~\ref{sec:MKBSC} is the central section of this paper, in 
which we define the MKBSC expansion, study its properties with respect to 
the preservation of certain classes of objectives, and describe 
how the construction can be used for the synthesis of first-order
knowledge-based strategies (the intensional view), or alternatively, 
of finite-memory strategies in the form of transducers (the dual,
extensional view). 
In Section~\ref{sec:IteratedMKBSC} we study the iterated MKBSC 
construction and how it can be used for the synthesis of 
higher-order knowledge-based strategies. 
In Section~\ref{sec:IMKBSC-Follow-up} we discuss the phenomenon of 
possible stabilisation in the iterated MKBSC construction, its 
implications on the strategy synthesis problem, and some limitations 
of the construction. 
Section~\ref{sec:related-work} discusses in some detail related work, while in 
Section~\ref{sec:Conclusion} we summarise our conclusions from the 
current work and provide directions for future work.

\section{Synthesis of knowledge-based strategies} 
\label{sec:Synthesis}

In this section, we offer an informal discussion on knowledge-based strategies and 
their synthesis. We then describe the concrete strategy profile synthesis problems studied in the paper. 

\subsection{Knowledge-based strategies}
\label{subsec:kbs}

By a \defstyle{knowledge-based strategy} we mean a strategy that uses suitably structured knowledge to determine the agent's course of action. That notion is conceptually very close to   \emph{knowledge-based programs}, cf.~\cite{DBLP:journals/dc/FaginHMV97,FHMV}, but here it is used  in the context of multi-agent games against Nature, defined in Section \ref{sec:MKBSC}. More precisely, a \defstyle{knowledge-based strategy} consists of:
\begin{enumerate}
\item a \defstyle{knowledge representation} (especially, for the dynamic knowledge) by a suitable data structure;

\item
a \defstyle{knowledge update} function that computes, after every transition in the game, the new knowledge state of the agent, from the old one, the action taken, and the observation made during and upon the transition;

\item
an \defstyle{action mapping}, from knowledge states to prescribed actions of the agent.
\end{enumerate}
 
The simplest knowledge-based strategies are the \emph{memoryless observation-based strategies}, cf. Section \ref{subsec:Strategies}, where the only knowledge used is the immediate observation of the agent on the current location. 
More generally, the agent's knowledge is represented by its full observation history, or by some finite abstraction of it. Thus, the most general and abstract case of knowledge-based strategies are the \emph{memory-based strategies}, where the used knowledge is not explicitly represented and structured but implicitly processed in the course of the play, by the strategy-computing device (e.g., the transducer\footnote{Also called ``(local) controller'' in the literature on planning and Dec-POMDP.}, in the case of finite-memory strategies). 
We call this approach to knowledge-based strategies ``extensional''. Alternatively, there is an ``intensional'' view, where the knowledge states do have structure, representing the dynamic knowledge of the agents during the course of a play. Several structures for knowledge representation, suitable for strategy design (though, some of them developed for the purpose of epistemic model checking, not strategy synthesis), have been studied, including: 
\emph{multi-agent epistemic models}~\cite{FHMV},  
\emph{knowledge structures}~\cite{Fagin1991,FHMV},  
\emph{$k$-trees}~\cite{DBLP:journals/iandc/Meyden98}, and
\emph{epistemic unfolding}~\cite{Berwanger2010,DBLP:conf/fsttcs/BerwangerKP11}. 
Some of the important questions arising here are: \emph{what knowledge is sufficient} to achieve a given objective, and  \emph{what is the minimal knowledge needed} for the purpose? 
We note two further related issues.

First,
structured knowledge in general requires memory to be stored and processed. Our idea of using structures for knowledge representation for strategy synthesis is to encode that knowledge in the states of the suitably expanded multi-agent games studied here, where memory-based strategies can be replaced by memoryless ones. 
The implication of this is that one can use as ``knowledge'' a data structure
that is simply a set of game locations, with the predefined
interpretation that it designates ``the most precise estimate an agent can
make about the current location, based on its initial knowledge about
the game and the history of all actions and observations made hitherto''
(and this is in effect what the expansion computes as locations of the
expanded game). 
Such knowledge can be updated after each action and
observation. 
Thus, we can interpret memoryless strategies in the
expanded games as knowledge-based strategies in the original games, with
this particular representation and interpretation of knowledge.

Second, when designing joint strategies of a team of agents acting towards a common goal, it can be essential to take into account their \emph{higher-order knowledge} about the other agents' knowledge. Intuitively, the reason for this is that a given agent from the coalition is not trying to achieve the objective on its
own (in which case it would have made sense for the agent to model the
other agents as ``nature''), but is collaborating with the rest. Therefore, 
a representation -- within an agent's knowledge -- of the estimates about
the current location, possibly made by the other agents, can offer higher
strategic ability for achieving joint objectives. The depth of such (nested) knowledge can increase without bound, and that generates a hierarchy of knowledge-representing structures and a respective hierarchy of knowledge-based strategies. Because that hierarchy may grow strictly, the search for a knowledge-based strategy for a given objective may never terminate, especially if such does not exist. This suggests that the strategy synthesis problem may generally be undecidable, and that is, indeed, the case~\cite{DBLP:conf/focs/PetersonR79}. 

\begin{figure}[ht]
    \centering
    \includegraphics[scale=.40]{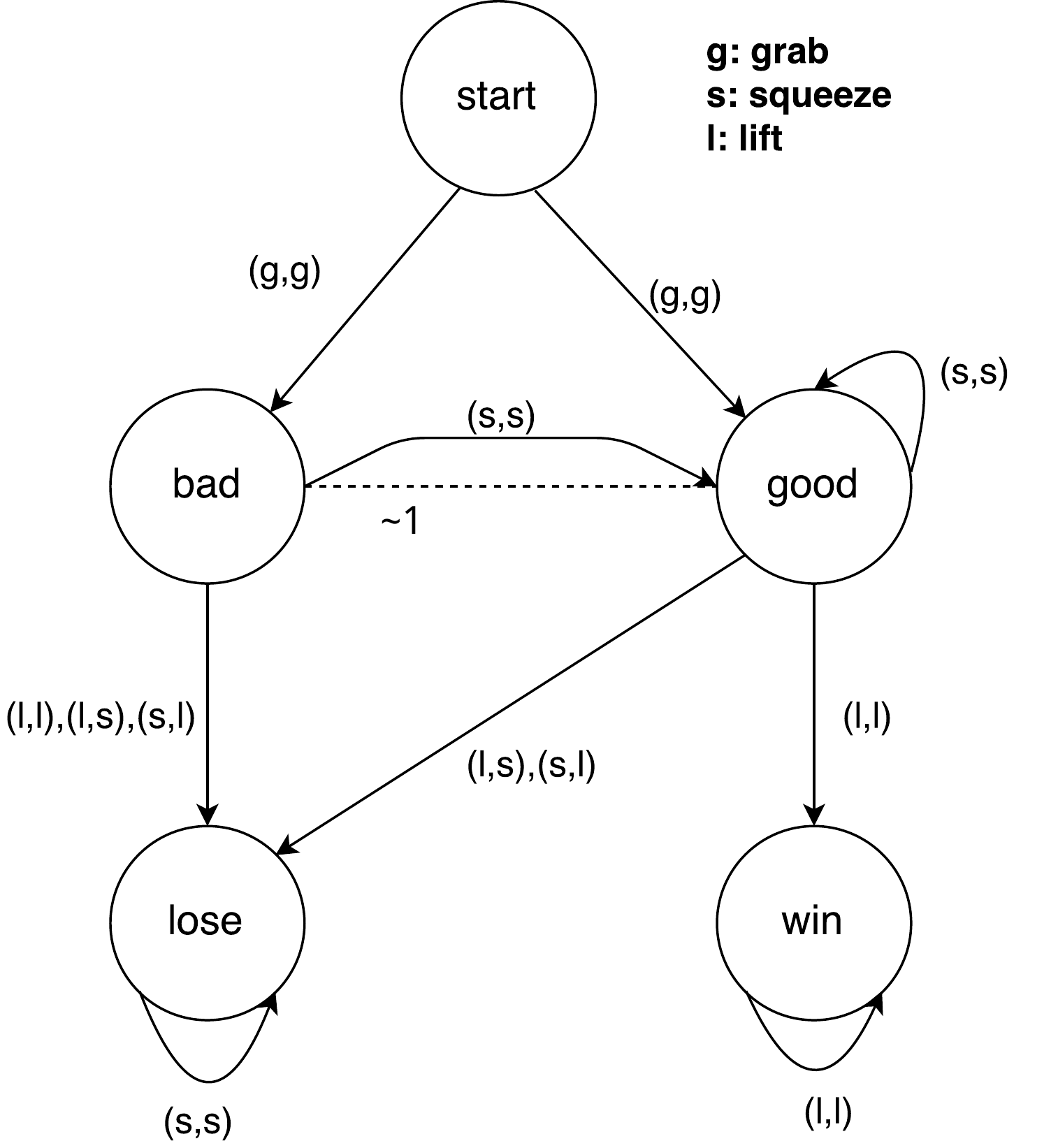}
    \caption{The two-robot cup-lifting game.}
    \label{fig:chem-accident}
\end{figure}

\begin{example}
\label{ex:robots}
Here is a running example, adapted from~\cite{lun-17-msc} and used further 
in the paper. 
Consider a scenario where two robots, henceforth referred to as robot~0 
and robot~1, must cooperate to lift a cup of acid. Both robots must first 
grab the cup. Grabbing the cup may non-deterministically result in a good 
overall grip or a bad one; the grip, however, can always be improved by 
simultaneously squeezing the cup. Then, both robots must simultaneously 
lift the cup; otherwise the cup will spill (and the game will be lost). 
In our scenario, robot~0 has a sensor that detects whether the overall
grip is good or not, while robot~1 does not have such a sensor (still, 
robot~1 can in some situations deduce from its actions and observations 
that the grip can only be a good one at this point). 

The scenario is modelled as the game depicted on 
Figure~\ref{fig:chem-accident}. 
While the formal notion of game will only be 
defined later, in Section~\ref{sec:MAGII}, the concrete game model should
be intuitively clear. The game is always in some location (i.e., node of the
graph), which changes as the result of the joint actions of the robots,
represented as action-pairs labelling the edges.
Only the available actions (at the respective location) and transitions that they enable are given in the figure. The uncertainty of robot~1 about the grip is modelled as a so-called observation; the corresponding indistinguishability equivalence~$\sim_1$ is depicted with a dotted line. 
In terms of the game graph, the objective of the two robots is to coordinate their actions so as to reach location $\mathsf{win}$. 

It should be easy to see that there cannot be an observation-based memoryless\footnote{Note that all memoryless strategies are also stationary.} 
 strategy that is winning. For a strategy to be observation-based, it has to respect the indistinguishability equivalences (i.e., observations) of the agents. Thus, robot~1 has to perform the same action in locations $\mathsf{bad}$ and $\mathsf{good}$. But for the team of robots to win, robot~1 needs to squeeze in location $\mathsf{bad}$ and lift in location $\mathsf{good}$.

The situation is similar when we consider first-order knowledge and the corresponding class of strategies. First-order knowledge of an agent is represented as a set of locations. The first-order knowledge of robot~0 
will always be a singleton set, representing the current location. The 
first-order knowledge of robot~1 will be, after the initialisation, 
represented as the set $\set{\mathsf{bad}, \mathsf{good}}$. Then, if after squeezing robot~1 makes the same observation (i.e., $\set{\mathsf{bad}, \mathsf{good}}$), it can now deduce that it only can be in $\set{\mathsf{good}}$. (Thus, knowledge is a refinement of what robots observe, using the power of deduction.) However, no first-order knowledge-based strategy can win the game, since knowing $\set{\mathsf{good}}$ is insufficient for robot~0 to decide whether to squeeze or to lift; to win the game, this must depend on whether robot~1 knows $\set{\mathsf{bad}, \mathsf{good}}$ or $\set{\mathsf{good}}$. But robot~0 has no knowledge of this when using first-order knowledge. 

This is remedied when using second-order knowledge: if robot~0 not only knows $\set{\mathsf{good}}$, but also knows whether the first-order knowledge of robot~1 is $\set{\mathsf{bad}, \mathsf{good}}$ or $\set{\mathsf{good}}$, it can make the correct decision whether to squeeze or to lift. As to robot~1, it will squeeze when knowing $\set{\mathsf{bad}, \mathsf{good}}$, and will lift when knowing $\set{\mathsf{good}}$. 
\end{example}

\noindent
We will use the above scenario as a running example and will elaborate on 
it throughout the paper.

Now that we have seen that higher-order knowledge can be necessary for 
achieving objectives, the question arises of how such knowledge can be 
computed and used for strategy synthesis. As already indicated, our 
approach is to apply suitable expansions that compute the higher-order 
knowledge of the agents, and then to search for memoryless strategies in 
the expanded games.

\subsection{Synthesis of knowledge-based strategies}
\label{subsec:KnowledgeBasedStrategies}

We study the problem of synthesis of knowledge-based strategies in the context of \emph{multi-agent games with imperfect information against Nature} (\magii, defined in Section \ref{sec:MAGII}): given such a game~$\G$ and an \defstyle{objective~$\obj$} (most generally, a given set of ``winning plays''),  
the task of a central authority -- the \defstyle{team supervisor} -- is to design a ``winning'' knowledge-based strategy profile (consisting of an individual knowledge-based strategy for each agent) for the team~$\Agt$ that guarantees the achievement of~$\obj$ regardless of the behaviour of the environment (Nature). We assume that the game structure is known by the supervisor, hereafter also called the  
\defstyle{(strategy) designer}, and is also common knowledge amongst all agents. 

Once the strategy profile is designed, each agent is assigned its strategy from the profile and the play begins. It is assumed that there is \emph{no explicit communication} between the agents during the play, that is, the only possible communication is by means of signalling through the agents actions  in the model, but not by explicit communicative actions, such as public or private announcements.
 
Then, four natural cases arise regarding the agents' knowledge and mutual observability during the play, that should be taken into account by the strategy designer \emph{in advance} and used for the design of their strategies:  

\begin{enumerate}
\item \textbf{Case (NN):} \emph{no strategy knowledge and no action observability}. The agents do not know the others' strategies\footnote{It is not trivial to formalise such knowledge, but what we intuitively mean by ``$\aga$ knowing the strategy of~$\agb$'' is that at every observation history for~$\aga$, that agent would only consider as possible those actions of~$\agb$ that are prescribed by the strategy of~$\agb$ known by~$\aga$ on some observation history that $\aga$ considers possible for~$\agb$ to have observed.} and cannot observe each others' actions during the play, but only their own. This will be our basic case of consideration, as we regard it as the most interesting one.  

\item \textbf{Case (NY):} \emph{no strategy knowledge but action observability}. The agents do not know the others' strategies,  but can observe each others' actions executed during the play. 

\item \textbf{Case (YN):} \emph{strategy knowledge but no action observability}. The agents know each others' strategies, i.e., the full strategy profile, but cannot observe the others' actions during the play. Note that, because of the partial observability, the agents generally do not know the other agents' observations, hence do not know exactly what actions they execute, so the case remains a priori non-trivial. 

\item \textbf{Case (YY):} \emph{strategy knowledge and action observability}. 
The agents know each others' strategies and can observe each others' actions 
during the play. 
For deterministic systems, this case is essentially reducible to a single-agent's play 
in a suitably modified model, as formally proved in \cite{DBLP:conf/prima/KazmierczakAJ14}. However, such reduction does not work explicitly in the case of non-deterministic games against Nature of the type we consider. Intuitively, this is because after the first transition neither of the agents may know the current state in the game, so even when knowing each others' strategies and observing each others' actions, they would not be able to fully coordinate their actions so as to be mergeable into a single agent. However, that can still be done in a suitably expanded model constructed as the product of the individual views by the different agents, by applying the general construction presented in Section 4.

\end{enumerate}

\noindent
Each of these cases has its own justification. For example, the supervisor may inform all agents about the full strategy profile, or may decide not to do that, for reasons of security or privacy. It is important to emphasize that the designer keeps in mind the specific case when designing the strategy profile, because these assumptions may make a difference for the existence of a winning strategy profile. It is generally clear that the more knowledge and observability the agents will have during the play, the greater the chance for existence of a winning strategy profile. What is not obvious, is whether all four cases are strictly different in that respect. 
To begin with, note that the actions observation ability generally helps agents to obtain more precise knowledge of the current location, and that can be used by the designer to construct a synchronised strategy profile. 
An illustrating example is given below.

\begin{example}
\label{ex:action-sync}
The figure below describes a simple turn-based \magii game with 2 agents
$\agone$ and $\agtwo$, whose collective objective is to reach one of the
$W$-states.

\begin{center}
\begin{tikzpicture}[->,>=stealth',shorten >=1pt,
node distance=2.0cm,
   el/.style = {inner sep=2pt, align=left},
                    thick]

\hspace{-5mm}

   \tikzstyle{every state}=[fill=white,draw=black,text=red]

   \node[state] (1)                    {$s_0 $};
   \node[state]         (2) [below  left of=1] {$ s_{1l} $};
   \node[state]         (3) [below right of=1] {$ s_{1r} $};
   \node[state]         (4) [below right of=2] {$ X $};
    \node[state]         (5) [below left of=2] {$ s_{2l} $};
    \node[state]         (6) [below right of=3] {$ s_{2r} $};
    \node[state]         (7) [below left of=5] {$ W $};
    \node[state]         (8) [below right of=5] {$ X $};
    \node[state]         (9) [below left of=6] {$ X $};
    \node[state]         (10) [below right of=6] {$ W $};

   \path (1) edge             node [el,above]  {$\mathsf{(*,*)}$ \ \ \ } (2)
             edge              node  [el,above] {\ \ \ \ $\mathsf{(*,*)}$} (3)
            (2) edge          node [el,below] {$\mathsf{(r,*)}$\ \ \ } (4)
             edge              node  [el,below]  {\ \ \ $\mathsf{(l,*)}$} (5)
           (3) edge 		node [el,below]  {\ \ \ $\mathsf{(l,*)}$}  (4)
              edge 			node [el,below]  {$\mathsf{(r,*)}$\ \ \ }  (6)
          (2) edge [-,dashed, bend right] node  [el,above]  {$\agtwo$} (3)
          (5) edge [-,dashed, bend right] node  [el,above] {$\agtwo$} (6)
         (5)    edge              node [el,below]  {\ \ \ $\mathsf{(*,l)}$}  (7)
             	edge              node [el,below] {$\mathsf{(*,r)}$\ \ \ }  (8)
         (6)    edge              node [el,below]  {\ \ \ $\mathsf{(*,l)}$}  (9)
             	edge              node [el,below] {$\mathsf{(*,r)}$\ \ \ } (10)       	
       (7) edge [loop below] node {$\mathsf{(*,*)}$}  (7)
       (8) edge [loop below] node {$\mathsf{(*,*)}$} (8)              
      (9) edge [loop below] node {$\mathsf{(*,*)}$} (9)
      (10) edge [loop below] node {$\mathsf{(*,*)}$} (10); 
      
\end{tikzpicture}
\end{center}

\noindent
The game goes as follows.
First, at $s_0 $ each agent has only one idling action, $\mathsf{*}$, and 
Nature decides to go left, to $ s_{1l} $, or right, to $ s_{1r} $. 
These successor states are only
distinguishable by~$\agone$ but not by~$\agtwo$ (indicated by a dotted
line in the diagram), who has only one action,  
$\mathsf{*}$, at each of these, whereas agent~$\agone$ gets to choose to 
go  left or right. If he does not match Nature's
choice the game ends in a bad state, denoted by~$X$, from which no
$W$-state is reachable. Otherwise, the game goes respectively in 
state~$s_{2l}$ or in~$s_{2r}$. These are, again, indistinguishable 
by~$\agtwo$, who is to make a left-right choice at each of them there, 
whereas $\agone$ has no choice (only one action, $\mathsf{*}$).    

The choice of~$\agtwo$ will be successful if and only if $\agtwo$ matches 
the choice of~$\agone$. 
Clearly, if $\agtwo$ could observe that action, she would easily succeed. However,
if $\agtwo$ cannot observe $\agone$'s action, there is no way that a
synchronised action profile can be pre-designed, because the correct
action of~$\agone$ cannot be decided in advance but only after Nature
has moved. This analysis applies regardless of whether the agents will
know each other's strategy at play time.
\end{example}

On the other hand, while knowing the other agents' strategies can be of importance for the knowledge update function of a knowledge-based strategy, it appears that it does not affect the \emph{existence} of a knowledge-based strategy profile achieving the team objective\footnote{This observation was first made to us by Dietmar Berwanger in a private communication.}. 
We consider this claim not intuitively obvious, but the reason, informally, is that the designer can use all the benefit from a common knowledge of the strategy profile at design time in order to synchronise all strategies, without having to provide the individual agents with that knowledge, as noted e.g. in \cite{DBLP:journals/jair/DibangoyeABC16}.

Likewise, in the \textbf{(YY)} case the designer has an apparently stronger power to synthesise a winning strategy profile than in each of the preceding cases. 
Still, the presumed common knowledge of the strategy profile at the play time is inessential for the existence and design of such a strategy profile, while the observability of the other agents' actions is (the above example also  distinguishes the cases \textbf{(YN)} and  \textbf{(YY)}); see 
also~\cite{DBLP:journals/jair/DibangoyeABC16}.

We are interested in synthesising knowledge-based strategies for each of the cases described above, but in this work we hereafter will focus mainly on the most challenging case \textbf{(NN}).

\section{Preliminaries on multi-agent games with imperfect information against Nature} 
\label{sec:MAGII}

We consider a fixed \defstyle{team} of~$n$  \defstyle{agents (players)}, 
$\Agt = \{\agone,...,\agn\}$, which aims to achieve a common goal. 

\begin{definition} 
A \defstyle{multi-agent game with imperfect information against Nature (\magii)} is a tuple $\G = (\Agt, \Loc, \init, \Act, \Delta, \Obs)$, 
where:
\begin{itemize}
\item[($i$)]   $\Loc$ is a set of \defstyle{locations}, usually assumed finite.

\item[($ii$)]  $\init \in \Loc$ is the \defstyle{initial location}.

\item[($iii$)] For each $\agi \in \Agt$, $\Act_\agi$ is a finite set of \defstyle{possible actions of agent $\agi$} (see remark 3 below).

\item[($iv$)]  $\Act = \Act_\agone \times \ldots \times \Act_\agn$ are the 
\defstyle{possible action profiles} (or joint actions)
of the team (see remark 3 below).

\item[($v$)]  $\Delta \subseteq \Loc \times \Act \times \Loc$ is a \defstyle{transition relation} between locations, with transitions labelled by action profiles.

\item[($vi$)]  For each $\agi \in \Agt$,  $\Obs_\agi$ is a partition of~$\Loc$, the blocks of which are the \defstyle{possible observations of agent~$\agi$}. 
Given any location~$\loc$, the unique observation for~$\agi$ containing~$\loc$ 
is denoted by $\obsi{\agi}{\loc}$. 
We denote with~$\sim_\agi$ the equivalence relation on locations induced by the partition.

\item[($vii$)]  $\Obs = \Obs_\agone \times \ldots \times \Obs_\agn$ 
is the set of all \defstyle{observation profiles} (or joint observations) 
of the team~$\Agt$. 
An observation profile $o \in \Obs$ is \defstyle{possible} 
iff $\cap_{\agi \in \Agt}\, o (\agi) \neq \varnothing$. 
We denote by $\Obs^p$ the set of possible observation profiles. 
\end{itemize}
\end{definition}

We already saw an example of a \magii in Example~\ref{ex:robots} in Section~\ref{subsec:kbs} above.
Some essential remarks are due here: 

\begin{enumerate}
\item 
The transition relation is assumed non-deterministic, in general, because the 
game is played against an unpredictable (and possibly stochastic) \defstyle{environment}, or \defstyle{Nature}, the possible behaviours of which are modelled through that non-determinism.
\item We study games of \emph{imperfect information}, where, in general, the agents can only partly observe the current location. This is modelled by  \emph{observational equivalence relations between locations} for each agent. Each such relation partitions the set of locations into blocks of indistinguishable locations, which are the possible observations of that agent. 
The particular case of perfect information is when all observations are singletons.
\item We implicitly assume that all actions of any given agent are available 
at every location. That is generally not justified, and we only assume it 
for the sake of technical convenience. 
Instead, we capture action availability via~$\Delta$: if an action 
profile~$\act$ contains an action that is not available for the respective 
agent at the given location $\loc$, then no transition is enabled by that 
action profile from~$\loc$, i.e., there is no~$\loc'$ such that 
$(l,\act, l') \in \Delta$. 
Furthermore, we assume that non-available actions will never be included 
in the designed strategy profiles.   
\item 
Lastly, a terminological remark: a \magii model is a variant of a ``factored model for a Qualitative Decentralized Partially Observable Markov Decision Problem (QDec-POMDP)'', as defined in \cite{DBLP:conf/aaai/SaffidineSZ18}, which is itself a variation of the ``QDec-POMDP model'' defined in \cite{DBLP:conf/aaai/BrafmanSZ13} (see further comments on these in Section 
\ref{sec:related-work}). The (not very essential) differences of our models and QDec-POMDP models are that: 

\begin{enumerate}
\item  we assume the agents' observations to be determined by the locations, rather than given by non-deterministic observation functions; 

\item the goal is not fixed in the model (e.g., as a reachability goal, as in the QDec-POMDP models) but is exogenously specified and can be reachability, safety, or more general, e.g., an LTL-definable objective; 

\item no horizon is explicitly specified in our models, and we implicitly assume it to be unbounded. 
\end{enumerate}

For a more detailed discussion on the relevant works on QDec-POMDPs and the relation of this study to them, see Section \ref{subsec:related-work-Dec-POMDP}. 

To avoid possible confusion, not to clutter further the terminology, and to emphasize the importance of \magii models on their own, we will not use the Dec-POMDP-based terminology but will adopt the acronym \magii throughout the paper. 
 
\end{enumerate}

\subsection{Plays and objectives} 
\label{subsec:plays}

The game on~$\G$ is played by the agents for infinitely many rounds (in general), starting from the initial location~$\init$. In each round, given the current location~$\loc \in \Loc$, each agent~$\agi$ chooses an action $a_\agi \in \Act_\agi$ that is available to $\agi$ at $\loc$, giving rise to an action profile~$\act \in \Act$. Then, Nature resolves the non-determinism by choosing the next location~$\loc' \in \Loc$ so that $(\loc, \act, \loc') \in \Delta$. 

A \defstyle{full play} in a \magii ~$\G$ is an infinite sequence $\fplay = \loc_0\sigma_1 \loc_1\sigma_2 \loc_2 \ldots$ of alternating locations and action profiles such that $\loc_0 = \init$ and  $\sigma_j \in \Act$ and $(\loc_j, \sigma_{j+1}, \loc_{j+1}) \in \Delta$ for all $j \geq 0$. 
A \defstyle{full history} is a finite prefix $\pi(j) = \loc_0 \sigma_1 \loc_1 \sigma_2  \ldots \loc_j$ of a full play~$\play$. 
A \defstyle{play} is the reduction of a full play to the subsequence of locations, $\play = \loc_0 \loc_1 \loc_2 \ldots$. 
Respectively, a \defstyle{history} is the reduction of a full history to the subsequence of locations, $\play(j) = \loc_0 \loc_1  \ldots \loc_j$. The last location on a history $\hist$ is denoted by $\last(\hist)$. 

From the perspective of any given agent, a play, resp.\ history, is a sequence of \emph{observations}, not of locations. Thus, for every agent $\agi$, a play $\play = \loc_0 \loc_1 \loc_2 \ldots$ generates an \defstyle{observation trace} of that play, which is the sequence  of respective observations $\obsi{\agi}{\loc_0}\, \obsi{\agi}{\loc_1}\, \obsi{\agi}{\loc_2} \ldots$  for that agent. Likewise, for any history $\hist$ we define the  \defstyle{observation history} for the given agent, being the respective finite prefix of the observation trace generated by the play containing the history.   

An \defstyle{objective} for the team $\Agt$ in the \magii~$\G$ is, most generally,  
a set of plays, declared as \defstyle{winning plays}  for $\Agt$. Often, an objective is  
expressed by a linear-time temporal logic formula~$\obj$, in the sense that the winning plays are precisely those satisfying that formula, where the atomic propositions in $\obj$ are assumed to have fixed interpretations in~$\G$.
Here are the most common types of objectives that we consider in this work: 
\begin{itemize}
\item 
A \defstyle{reachability objective} can be defined by a non-empty set of locations 
$\reach \subseteq \Loc$. 
A play $\play = \loc_0 \loc_1 \loc_2 \ldots$ is winning if it visits some location in~$\reach$, i.e., if $\loc_i \in \reach$ for some $i \geq 0$.

A reachability objective is \defstyle{observable for an agent~$\agi$}, if it 
is a union of observations for~$\agi$, and can therefore be defined alternatively 
as a set $\reach \subseteq \Obs_\agi$ of observations for~$\agi$; 
the objective is \defstyle{observable (for the team)} if it is observable, 
at some point in time, for at least one agent in the team, 
and thus $\reach \subseteq \cup_{\agi \in \Agt}\, \Obs_\agi$.
The latter notion of observability for the team may not always be
justified, and indeed alternative formulations are also possible.
However, our results on strategy preservation (see 
Section~\ref{subsec:strat-pres} below) are for the notion stated here. 

\item 
A \defstyle{safety objective} is defined by a non-empty set of locations $\safe \subseteq \Loc$. 
A play $\pi = \loc_0 \loc_1 \loc_2 \ldots$ is winning if it only visits  
locations in~$\safe$, i.e., if $\loc_i \in \safe$ for all $i \geq 0$.

Observable safety objectives are defined similarly to observable reachability objectives. Thus, an observable (for the team) safety objective is a set
$\safe \subseteq \cup_{\agi \in \Agt}\, \Obs_\agi$, and to win, at every
point in time at least one agent must observe the objective. 
Again, alternative formulations are possible, but our results
are for the given one. 
\end{itemize}
In this work we will be concerned with reachability and safety objectives 
that are observable for the team. 

\subsection{Observation-based strategies} 
\label{subsec:Strategies}

In games with imperfect information the simplest type of agents' strategies are based on agents' observations. These are also the simplest type of \emph{knowledge-based strategies}, more generally discussed in Section \ref{subsec:kbs}. 

Given a \magii $\G$, a \defstyle{(deterministic) perfect-recall observation-based strategy} for an agent $\agi$ is a mapping $\strat_\agi :  \Obs^+_\agi \rightarrow \Act_\agi$ prescribing for every observation history $\hist$ for the agent $\agi$ an action $\strat_\agi(\hist)$ that is available for $\agi$ at $\last(\hist)$. 

An observation-based strategy for $\agi$ is called \defstyle{memoryless} (or \defstyle{positional}) if it only takes into account the \emph{current} observation (the last one of the observation history). Such a strategy can be simply presented as a mapping of the type $\Obs_\agi \rightarrow \Act_\agi$.

A \defstyle{finite-memory observation-based strategy} is commonly modelled as a finite-state \textbf{transducer}, or \textbf{Moore machine}, reading game histories and mapping them to actions by using memory states, and is formally defined as follows. 

\begin{definition}[Finite-Memory Strategy]
\label{def:finite-memory-strategy}
A \defstyle{finite-memory observation-based strategy} for agent~$\agi$ in a \magii 
$\G = (\Agt, \Loc, \init, \Act, \Delta, \Obs)$
is a structure  
$\moore_\agi = (M, m_0, \Obs_\agi, \Act_\agi, \tau, \gamma)$, where:
\begin{itemize}
\item[($i$)]   $M$ is a finite set of \defstyle{memory states}; 

\item[($ii$)]  $m_0 \in M$ is the \defstyle{initial memory state};

\item[($iii$)]  $\tau : M \times \Obs_\agi \rightharpoonup M$ is a (partial) transition function; 

\item[($iv$)] $\gamma : M \rightarrow \Act_\agi$ is a mapping from memory states to actions for~$\agi$. 
\end{itemize}
Agent~$\agi$ follows the strategy encoded by $\moore_\agi$ as follows. In each round, ~$\agi$ selects as its next action $\gamma (m)$, where $m$ is the current memory state of $\moore_\agi$ (initially~$m_0$). After the team has applied its action profile  and Nature has chosen the next location~$\loc$, agent~$\agi$ makes the corresponding observation $\obsi{\agi}{\loc}$ and  updates its memory state to $\tau (m, \obsi{\agi}{\loc})$. 
\end{definition}
Note that~$\tau$ can be partial, since, in the 
context of a given \magii~$\G$, some combinations of memory states and 
observations might never occur during play. 

A  \defstyle{perfect-recall} (resp.,  \defstyle{positional}, or  \defstyle{finite-memory})  \defstyle{observation-based strategy profile} is a strategy profile consisting of {perfect-recall} (resp., {positional}, or {finite-memory}) observation-based strategies. 

An \textbf{outcome} of an agent's (observation-based) strategy is any play in which the agent chooses its actions according to that strategy. Likewise we define an \textbf{outcome of a strategy profile}. Note that, because of potential non-determinism, such outcomes are generally not unique.  
A strategy profile is  \textbf{winning for an objective $\obj$} if all of its outcomes belong to $\obj$. 

It should be noted that the restriction to observable objectives can be 
partly overcome by using the following technical trick. Given a \magii~$\G$, 
one can transform~$\G$ to another game~$\G'$ by adding to the set 
of agents a new, ``dummy'' agent whose observations are singletons,
and who has in all its locations a single idling action at its disposal 
that is appended to all existing action profiles. 
Clearly, $\G$ and $\G'$ have the same strategies; furthermore, according 
to our definition, all objectives in~$\G'$ are observable, even the ones
that correspond to objectives in~$\G$ that are not observable in~$\G$. 
Then, if there is no winning strategy in~$\G'$ for a given objective, 
there is no winning strategy in~$\G$ either. 
On the other hand, if there is a winning strategy in~$\G'$, it may not be 
directly usable if the objective is only observed by the dummy agent, 
since the latter does not really exist in the actual game~$\G$.

\subsection{Knowledge-based strategies} 
\label{subsec:kb-strat}

We now present a framework for defining knowledge-based strategies of an 
agent in a \magii game. It consists of two parts: \\

I. An \defstyle{information update module for the agent~$\agi$ in the game~$\G$},
where: 
\begin{itemize}
\item $\Kn{\agi}$ represents the a priori knowledge (information) of the agent $\agi$ about the game, the other agents, their strategies, etc. (to be specified). 

\item $\Inf{\agi}$ is the set of possible knowledge states of the agent $\agi$. 

\item $\Act_\agi$ is the set of actions that $\agi$ can take. 

\item $\Obs_\agi$ is the set of observations that $\agi$  can make 
(about locations and actions).
\end{itemize}

The information update module is defined as a mapping:
\[
\upd{\G}{\agi}{\Kn{\agi}}:  \Inf{\agi} \times \Act_\agi  \times  \Obs_\agi \to \Inf{\agi}
\]

\medskip 
II. An \defstyle{information (or, knowledge) based strategy} for $\agi$: a mapping: 
\[
\str{\agi}:   \Inf{\agi}  \to \Act_\agi
\]
Observe that finite-memory observation-based strategies as defined
in Definition~\ref{def:finite-memory-strategy} are an instance of 
the above framework, with $M$ for $\Inf{\agi}$, $\gamma$ for $\str{\agi}$,
and $\tau$ for $\upd{\G}{\agi}{\Kn{\agi}}$, where the latter is 
restricted to actions as prescribed by $\str{\agi}$.

The rationale behind the above formulation is the following.
The possible knowledge states are abstractions 
over the sequences of actions and observations of the agents,
and are updated upon each action and observation. In a
knowledge-based strategy, the next action of an agent is
completely determined by its current knowledge state.

\section{A multi-agent knowledge-based subset construction}
\label{sec:MKBSC}

Here we introduce and study a new construction, which generalises to the multi-agent case the well-known \emph{knowledge-based subset construction}  (KBSC)~\cite{DBLP:journals/jcss/Reif84}, which transforms single-agent games with imperfect information to (expanded) single-agent games with perfect information. The KBSC is strategy-preserving for the large class of parity objectives, cf.~\cite{DBLP:journals/lmcs/RaskinCDH07}. We do not present here the construction on its own, but as a component of the generalised construction. 

Note that the results of this section concern first-order knowledge only, whereas higher-order knowledge will be studied in the following section, in the context of the iterated construction.

\subsection{Generalising the KBSC}
\label{subsec:gen-kbsc}

To connect knowledge-based strategies in multi-agent games to observation-based memoryless strategies in expanded games, we propose a generic scheme for expansion, which is independent of the concrete knowledge representation and the concrete assumptions on what the agents can know and observe.

Our generic scheme for extending the KBSC to the multi-agent setting consists of four stages:  

\begin{enumerate}
\item 
\textbf{Projection}: for each agent~$\agi \in \Agt$, compute the individual views of the input game~$\G$, based on what the agent knows, does and sees. This stage results in $n$ single-agent games with imperfect information.
\item
\textbf{Expansion}: expand each of the individual views with the KBSC. The results are $n$ single-agent games with perfect information.
\item
\textbf{Composition}: combine the individual expansions by using a product construction, resulting in a single multi-agent game with perfect information.
\item
\textbf{Partition}: define each agent's observations as induced by the composition product, reflecting their local knowledge.
The final result is a multi-agent game with imperfect information.
\end{enumerate}

A concrete instantiation of that scheme is the \defstyle{Multi-Agent Knowledge-Based Subset Construction (MKBSC)} for the case \textbf{(NN)}, defined below. An implementation of the MKBSC as a tool\footnote{Available from \texttt{github.com/helmernylen/mkbsc}.} is described in~\cite{nyl-jac-18-bsc}. The game graphs in the rest of the paper have been produced and visualised with this tool.
 
\begin{definition}[MKBSC]
\label{def:mkbsc}
Let~$\G = (\Agt, \Loc, \init, \Act, \Delta, \Obs)$ be a \magii.
\begin{enumerate}
\item
\textbf{Projection}: 
Given an agent $\agi \in \Agt$, we define the projection of~$\G$ onto~$\agi$ 
as the single-agent game with imperfect information: 
\[\G|_\agi \defeq (\Loc, \init, \Act_\agi, \Delta_\agi, \Obs_\agi),\] 
where 
$(l, \act_\agi, l') \in \Delta_\agi$ \ iff \ 
there exists
$\act \in \Act$ such that $\act(\agi) = \act_\agi$ and $(l, \act, l') \in \Delta$.

\item
\textbf{Expansion}: Given $\G|_\agi$ as above, we define its 
KBSC expansion, following~\cite{DBLP:journals/lmcs/RaskinCDH07}, 
as the single-agent game with perfect information:
\[(\G|_\agi)^\mathsf{K} \defeq (S_\agi, s_{I,\agi}, \Act_\agi, \Delta^\mathsf{K}_\agi),\]
where 
$S_\agi \defeq \setdef{s \in 2^\Loc \backslash \set{\varnothing}}{\exists o_\agi \in \Obs_\agi.\ s \subseteq o_\agi}$ 
is the set of possible knowledge states of agent~$\agi$,
$s_{I,\agi} \defeq \set{\init}$ is its initial knowledge state, and 
$\Delta^\mathsf{K}_\agi \ \defeq$ \\ 
$\big\{(s, \act_\agi, s') \in S_\agi \times \Act_\agi \times S_\agi \mid 
\exists o_\agi \in \Obs_\agi.\  s' = \setdef{l' \in o_\agi}{\exists l \in s.\ (l, \act_\agi, l') \in \Delta_\agi}\big\}$.

\item
\textbf{Composition}: Given $(\G|_\agi)^\mathsf{K}$ as above, 
for all $\agi \in \Agt$, we construct their synchronous product, 
with \emph{joint knowledge states} $S \defeq \times_{\agi \in \Agt} S_\agi$ 
and transitions~$\Delta^\mathsf{K}$ labelled by joint actions~$\Act$.
The initial knowledge state $s_I \in S$ is the tuple 
$(s_{I,\agi})_{\agi \in \Agt}$.
We \emph{prune} the product by removing inconsistent knowledge 
states~$s$, i.e., tuples of sets of locations, the intersection 
$\cap_{\agi \in \Agt}\, s (\agi)$ 
of which is empty, and unrealisable transitions, i.e., transitions
$s \stackrel{\act}{\longrightarrow} s'$ for which there is no
transition $l \stackrel{\act}{\longrightarrow} l'$ in~$\Delta$
such that $l \in \cap_{\agi \in \Agt}\, s (\agi)$ and 
$l' \in \cap_{\agi \in \Agt}\, s' (\agi)$.

\item
\textbf{Partition}: We define the observations $\Obs^\mathsf{K}_\agi$ 
of every agent $\agi \in \Agt$ as induced by their local knowledge:
$$ (s_1, s_2) \in \Obs^\mathsf{K}_\agi \>\defequiv\> s_1 (\agi) = s_2 (\agi) $$
The observations thus represent indistinguishability with respect to 
knowledge rather than locations. 
\end{enumerate}
\end{definition}
\noindent
The result is the \defstyle{MKBSC expansion}: 
\[\G^\mathsf{K} = (\Agt, S, s_I, \Act, \Delta^\mathsf{K}, \Obs^\mathsf{K})\]
of~$\G$, which is a \magii. 
Since only the part reachable from~$s_I$ is of interest, the rest is 
disregarded.

A  different, but equivalent formulation was originally proposed 
in~\cite{lun-17-msc} by the third co-author. The formulation given here makes explicit how the resulting game is composed of individual expansions, which is the basis for several of our results below.

\begin{figure}[ht]
    \centering
    \includegraphics[scale=.4]{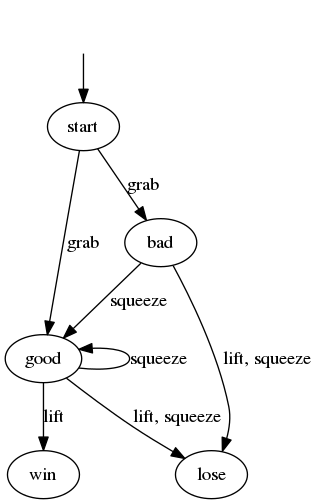}
    \hspace{10truemm}
    \includegraphics[scale=.4]{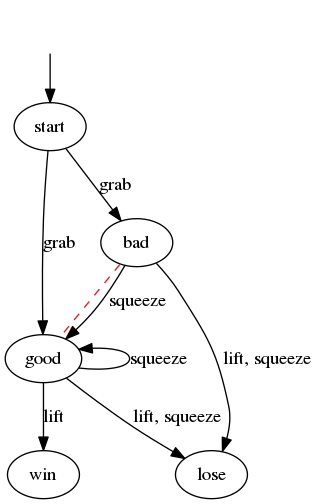}
    \caption{The individual projections $\G|_0$ and~$\G|_1$.}
    \label{fig:indiv-proj}
\end{figure}

\begin{figure}[ht]
    \centering
    \includegraphics[scale=.4]{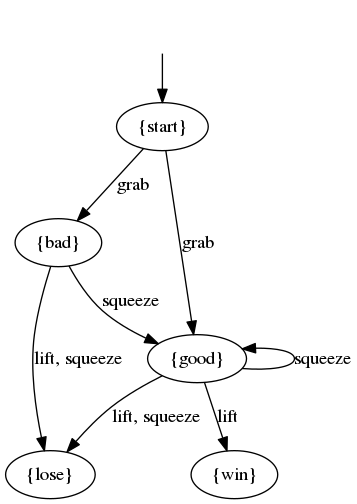}
    \hspace{10truemm}
    \includegraphics[scale=.4]{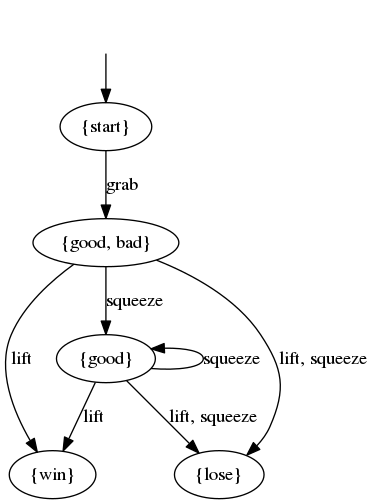}
    \caption{The individual expansions $(\G|_0)^\mathsf{K}$ 
             and~$(\G|_1)^\mathsf{K}$.}
    \label{fig:indiv-exp}
\end{figure}

\begin{figure}[ht]
    \centering
    \includegraphics[scale=.35]{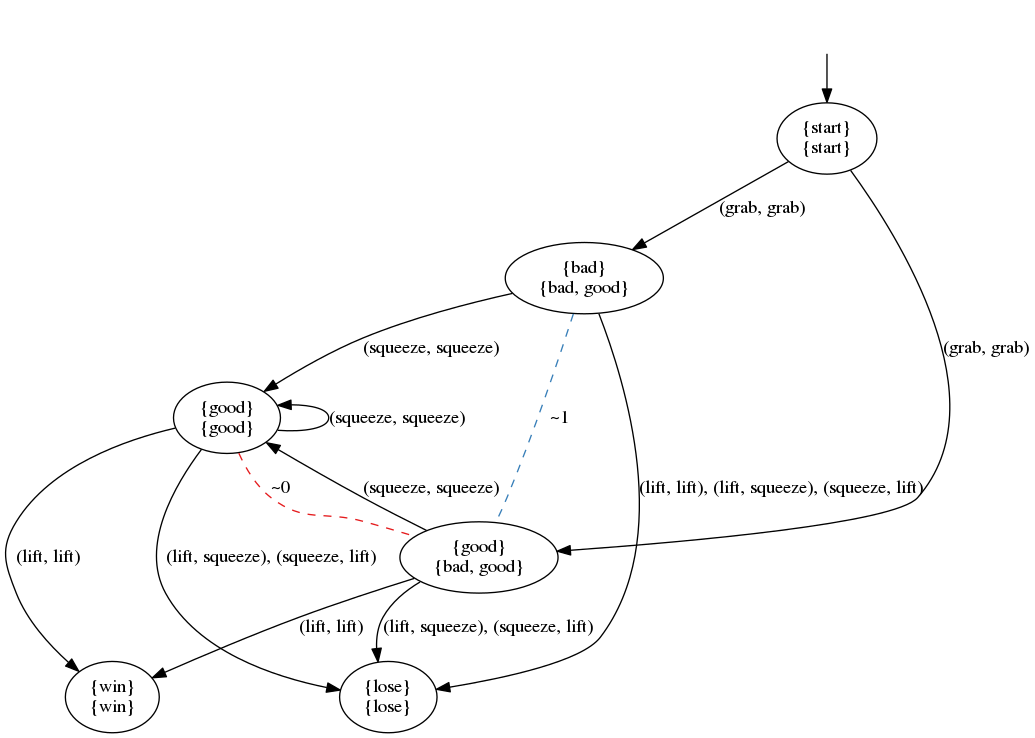}
    \caption{The pruned product $\G^\mathsf{K}$.}
    \label{fig:prod}
\end{figure}

\begin{example}
\label{ex:mkbsc} 
We illustrate the MKBSC construction on our running Example~\ref{ex:robots},
here to be denoted as~$\G$.
The individual projections $\G|_0$ and~$\G|_1$ are shown in Figure~\ref{fig:indiv-proj},
while the individual expansions $(\G|_0)^\mathsf{K}$ and~$(\G|_1)^\mathsf{K}$
of $\G|_0$ and~$\G|_1$, respectively, are shown in Figure~\ref{fig:indiv-exp}.
And finally, the pruned product~$\G^\mathsf{K}$ of $(\G|_0)^\mathsf{K}$ 
and~$(\G|_1)^\mathsf{K}$ is shown in Figure~\ref{fig:prod}.

Due to the pruning, there only are consistent knowledge states 
in~$\G^\mathsf{K}$. 
There is an edge labelled $(\mathsf{squeeze}, \mathsf{squeeze})$ from
vertex $(\set{\mathsf{bad}}, \set{\mathsf{bad}, \mathsf{good}})$ 
to vertex $(\set{\mathsf{good}}, \set{\mathsf{good}})$ in~$\G^\mathsf{K}$ 
because there is an edge labelled $\mathsf{squeeze}$ from vertex 
$\set{\mathsf{bad}}$ to vertex $\set{\mathsf{good}}$ in~$(\G|_0)^\mathsf{K}$,
an edge labelled $\mathsf{squeeze}$ from  
$\set{\mathsf{bad}, \mathsf{good}}$ to  
$\set{\mathsf{good}}$ in~$(\G|_1)^\mathsf{K}$,
and an edge labelled $(\mathsf{squeeze}, \mathsf{squeeze})$
from location $\mathsf{bad}$ to location $\mathsf{good}$ in~$\G$. 
Note that if the latter edge had not been present in~$\G$, the 
discussed edge in~$\G^\mathsf{K}$ would have been unrealisable,
and would have been pruned out. 

\end{example}

While in this paper we do not focus on the algorithmic aspects of the 
MKBSC construction, a naive analysis of its space complexity reveals 
that: (a)~projection preserves the locations of~$\G$, (b)~expansion 
(being a subset construction) is worst-case exponential in the number 
of locations of~$\G$, and (c)~composition results in the product of 
the numbers of locations of the individual expansions.
The space complexity of the construction can thus be upper-bounded by 
$O (2^{|\Loc| \cdot |\Agt|})$. 

\noindent
For the \textbf{(NY)} case, the MKBSC construction can be adapted 
as follows:
\begin{enumerate}
\setlength{\itemsep}{-3pt}
\item
The projections do not filter out the complementary actions of the other
agents, but keep the full joint actions, and only abstract from the
observations of the other agents; this results in the games
$\G|_\agi \defeq (\Loc, \init, \Act, \Delta, \Obs_\agi)$. 
\item 
The individual expansion stage  remains  unchanged.
\item
The synchronous product now has to synchronise on common joint actions.
\item
The partition stage  remains unchanged.
\end{enumerate}
For example, consider the game~$\G$ shown in Figure~\ref{fig:strat-loss} left.
In the \textbf{(NY)} case, its expansion~$\G^\mathsf{K}$ is isomorphic
to the original game~$\G$, but has knowledge states $(\set{l}, \set{l})$
for the corresponding locations~$l$ in~$\G$. 

In the sequel, unless otherwise specified, we only refer to the 
\textbf{(NN)} case.

\subsection{Strategy preservation}
\label{subsec:strat-pres}

In this section we present results on the preservation of \emph{observation-based perfect recall strategies} for observable reachability objectives. 
Note that every observable reachability objective
$\reach \subseteq \cup_{\agi \in \Agt}\, \Obs_\agi$ 
in a \magii~$\G$ with observations~$\Obs$ translates uniformly to an observable 
reachability objective $\reach^\mathsf{K}$ in~$\G^\mathsf{K}$, as follows:  
$$\reach^\mathsf{K} \defeq \setdef{s \in S}{\exists \agi \in \Agt.\ \exists o \in \reach \cap \Obs_\agi.\ s (\agi) \subseteq o}$$
Likewise, every observable safety objective~$\safe$ in~$\G$ translates uniformly to an observable safety objective $\safe^\mathsf{K}$ in~$\G^\mathsf{K}$. 

The synchronous product of the MKBSC construction is a form of
\emph{existential abstraction}, and can give rise to ``spurious'' plays 
in~$\G^\mathsf{K}$ that are not present in~$\G$. Such spurious plays
can give rise to spurious outcomes of a given strategy, and can thus
prevent a strategy from achieving a given objective. 

\begin{figure}[ht]
    \centering
    \includegraphics[scale=.48]{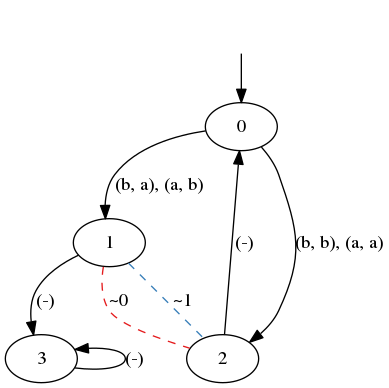}
    \hspace{7truemm}
    \includegraphics[scale=.4]{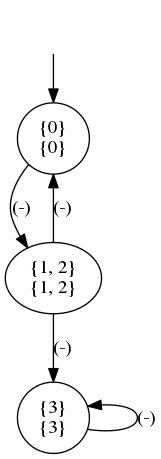}
    \caption{A game $\G$ and its expansion $\G^\mathsf{K}$.}
    \label{fig:strat-loss}
\end{figure}

\begin{example}
Consider the two-agent game~$\G$ and its expansion~$\G^\mathsf{K}$,
shown in Figure~\ref{fig:strat-loss} 
(where the symbol ``(-)'' is used to denote any joint action), 
and let $\reach \defeq \set{\set{3}}$ be
our observable reachability objective. The game is easily won in~$\G$ with the
profile of observation-based memoryless strategies, where the first agent
always does action~$a$, while the second agent always does~$b$. In
the game~$\G^\mathsf{K}$, however, the (corresponding) strategy is thwarted 
by the outcome 
$(\set{0}, \set{0})\ (\set{1, 2}, \set{1, 2})\ (\set{0}, \set{0})$,
which is spurious: there is no corresponding outcome $0\, 1\, 0$ or 
$0\, 2\, 0$ in~$\G$. 
\end{example} 

Thus, it is not possible to preserve arbitrary winning strategies from~$\G$ 
to~$\G^\mathsf{K}$. 
However, under the additional condition that for every joint knowledge 
state~$s$ that is reachable from the initial one in~$\G^\mathsf{K}$, 
$\cap_{\agi \in \Agt}\, s (\agi)$ is a singleton set, no spurious plays 
exist. 
This condition can be stated as  \defstyle{perfect distributed 
knowledge} (or PDK for short\footnote{In the literature 
on Dec-POMDP this condition is also called 
``joint observability'' \cite{DBLP:series/sbis/OliehoekA16}, 
``collective observability'' \cite{DBLP:journals/aamas/SeukenZ08} 
and 
``decentralised full observability'' \cite{DBLP:conf/cdc/AmatoCGUK13}, and the 
models satisfying it are called 
``decentralized Markov decision processes (Dec-MDP)''. 
}. 

For instance, the game~$\G^\mathsf{K}$ from Figure~\ref{fig:prod} satisfies 
the PDK, while $\G^\mathsf{K}$ from Figure~\ref{fig:strat-loss} does not.
An obvious sufficient condition on~$\G$ to guarantee that $\G^\mathsf{K}$
fulfills the PDK condition is that no two distinct locations of~$\G$ are
indistinguishable by all agents (or equivalently, that any two distinct 
locations of~$\G$ are distinguishable by at least one 
agent)\footnote{%
Therefore, the technical trick of adding a dummy agent described at the 
end of Section~\ref{subsec:Strategies} enforces the PDK condition.}.

When game~$\G^\mathsf{K}$ satisfies the PDK, one can view~$\G^\mathsf{K}$
as a \emph{refinement} of the original game~$\G$, in the sense that the
locations in~$\G^\mathsf{K}$ are obtained from ``splitting'' locations
of~$\G$. Since any two locations of~$\G^\mathsf{K}$ that derive from
the same location in~$\G$ will always be distinguishable by at least one
agent, one can say that, while the MKBSC does not necessarily eliminate 
imperfect information (as it does in the single-agent case), it in some
sense decreases the degree of imperfectness. 

The following result generalises Lemma~3.1 from~\cite{DBLP:journals/lmcs/RaskinCDH07}.

\begin{lemma}
\label{lem:MKBSCCharacterisation}
Let $\G = (\Agt, \Loc, \init, \Act, \Delta, \Obs)$ be a \magii for a set 
of agents~$\Agt$, and let 
$\G^\mathsf{K} = (\Agt, S, s_I, \Act, \Delta^\mathsf{K}, \Obs^\mathsf{K})$ 
be its MKBSC expansion. 
Further, let $s \in S$, $\act \in \Act$ and $o \in \Obs^p$.
Define the set:
$$ X \>\defeq\> \setdef{l' \in \cap_{\agi \in \Agt}\,  o (\agi)}
  {\exists l \in \cap_{\agi \in \Agt}\, s (\agi).\ (l, \act, l') \in \Delta} $$
Then, $X$ is non-empty if and only if there is $s' \in S$ such 
that $(s, \act, s') \in \Delta^\mathsf{K}$ and for all $\agi \in \Agt$, 
$s' (\agi) \subseteq o (\agi)$.
This $s' \in S$ is then unique and we have
$X \subseteq \cap_{\agi \in \Agt}\, s' (\agi)$,
and if $\G^\mathsf{K}$ fulfills the PDK condition, 
we have $X = \cap_{\agi \in \Agt}\, s' (\agi)$.
\end{lemma}

\begin{proof}
Let $s \in S$, $\act \in \Act$ and $o \in \Obs^p$. The stated equivalence 
is a direct consequence of the expansion and composition steps of Definition~\ref{def:mkbsc}; in particular, when there is $s' \in S$ with 
the stated properties,
the set~$X$ cannot be empty since otherwise the transition~$(s, \act, s')$ 
would be pruned out (as being unrealisable) in the composition step. 
Furthermore, if $\G^\mathsf{K}$ fulfills the PDK condition, the 
set $\cap_{\agi \in \Agt}\, s' (\agi)$ must be a singleton, and
the two sets must therefore be equal.
\end{proof}

This result should apply likewise to stochastic, rather than non-deterministic models, of the type studied in the literature on Dec-POMDP.

The result lifts naturally to observation histories: every sequence~$\pi$ 
of joint actions and joint observations of~$\G$ (where~$\pi$ is not
necessarily a path in~$\G$) gives rise to at most 
one path in~$\G^\mathsf{K}$ such that at each corresponding step of the 
two sequences, $s(\agi) \subseteq o(\agi)$ holds for all $\agi \in \Agt$. 
Furthermore, every full play~$\pi$ in~$\G$ gives rise to exactly one 
full play in~$\G^\mathsf{K}$ that is consistent with the actions and 
the observations of the agents. 

We obtain the following result on strategy preservation under the 
PDK condition. 

\begin{theorem}[Strategy Preservation]
\label{thm:strat-pres}
Let~$\G$ be a \magii, and let $\G^\mathsf{K}$ be its MKBSC expansion.
Assume that $\G^\mathsf{K}$ fulfills the PDK condition. 
Let~$\reach$ be an observable reachability objective in~$\G$, and 
$\reach^\mathsf{K}$ be its translation for~$\G^\mathsf{K}$. 
If there is a winning profile of observation-based perfect recall 
strategies in~$\G$ for~$\reach$, then there is also one in~$G^\mathsf{K}$ 
for~$\reach^\mathsf{K}$. 
\end{theorem}

\begin{proof}
Informally, given a profile~$\set{\alpha_\agi}_{\agi \in \Agt}$ 
of observation-based perfect recall strategies in~$\G$, every 
agent~$\agi \in \Agt$ plays in~$\G^\mathsf{K}$ by: 
\begin{enumerate}
\setlength{\itemsep}{-3pt}
\item 
recording the history of individual observations in~$\G^\mathsf{K}$ it 
has made so far during the play, 
\item
converting this sequence to the corresponding sequence of 
observations of the agent in~$\G$, and 
\item
taking the action prescribed by~$\alpha_\agi$ for that 
sequence of observations. 
\end{enumerate}

Formally, let $\set{\alpha_\agi}_{\agi \in \Agt}$, 
where $\alpha_\agi : \Obs_\agi^+ \rightarrow \Act_\agi$ for all~$\agi \in \Agt$, 
be a winning profile of observation-based perfect recall strategies in~$\G$ 
for the observable reachability objective~$\reach$. 
For all $\agi \in \Agt$, define the functions 
$\alpha_\agi^\mathsf{K} : (\Obs^\mathsf{K}_\agi)^+ \rightarrow \Act_\agi$ 
as the function compositions 
$\alpha_\agi^\mathsf{K} \defeq \alpha_\agi \circ \mathsf{ob}_\agi$, where 
for every $o^\mathsf{K}_\agi \in \Obs^\mathsf{K}_\agi$, 
$\obi{\agi}{o^\mathsf{K}_\agi}$ denotes the unique observation 
$o_\agi \in \Obs_\agi$ in~$\G$ such that $A \subseteq o_\agi$
holds for the common $\agi$'th component~$A$ of the tuples 
comprising~$o^\mathsf{K}_\agi$,
and where $\mathsf{ob}_\agi$ is then lifted to sequences. 
Thus the functions define a profile 
$\set{\alpha_\agi^\mathsf{K}}_{\agi \in \Agt}$
of observation-based perfect recall strategies in~$G^\mathsf{K}$. 
We show that this profile is winning in~$G^\mathsf{K}$ for the 
objective~$\reach^\mathsf{K}$ whenever $G^\mathsf{K}$ satisfies the 
PDK condition.

Let $\pi^\mathsf{K} = s_0 \sigma_0 s_1 \sigma_1 \ldots$ be an arbitrary outcome 
of $\set{\alpha_\agi^\mathsf{K}}_{\agi \in \Agt}$ in~$\G^\mathsf{K}$.
Then, by Lemma~\ref{lem:MKBSCCharacterisation} and since $G^\mathsf{K}$ 
satisfies the PDK condition, the sequence
$\pi = l_0 \sigma_0 l_1 \sigma_1 \ldots$ such that 
$\set{l_k} = \cap_{\agi \in \Agt}\, s_k (\agi)$ for all $k \geq 0$, 
must be a full play  in~$\G$. 
But then $s_k (\agi) \subseteq \obsi{\agi}{l_k}$ for all $\agi \in \Agt$ 
and $k \geq 0$, and thus, by the definitions of $\alpha_\agi^\mathsf{K}$, 
$\pi$~must be an outcome of 
$\set{\alpha_\agi}_{\agi \in \Agt}$ in~$\G$, and must be winning for~$\reach$, 
i.e., there is an agent $\agi \in \Agt$ and an index $k \geq 0$ such that
$\obsi{\agi}{l_k} \in \reach$. 
But then there is also  an agent $\agi \in \Agt$, 
an index $k \geq 0$ and an observation 
$o \in \reach \cap \Obs_\agi$ such that $s_k (\agi) \subseteq o$, 
and hence $\pi^\mathsf{K}$~is winning for~$\reach^\mathsf{K}$.
Since $\pi^\mathsf{K}$ is arbitrary, the profile
$\set{\alpha_\agi^\mathsf{K}}_{\agi \in \Agt}$ must be winning 
in~$G^\mathsf{K}$ for the objective~$\reach^\mathsf{K}$. 
\end{proof}

It is easy to see from the proof that winning profiles of observation-based 
perfect recall strategies for \emph{observable safety objectives}
$\safe \subseteq \cup_{\agi \in \Agt}\, \Obs_\agi$ 
are also preserved, by a slightly modified argument: 
in the proof, simply replace ``there is an index $k \geq 0$'' 
with ``for all indices $k \geq 0$''. 

It is important to observe that the above proof is constructive and 
shows how observation-based \emph{memoryless} strategies
$\alpha_\agi : \Obs_\agi \rightarrow \Act_\agi$ in~$\G$ 
are mapped to observation-based memoryless strategies 
$\alpha_\agi^\mathsf{K} \defeq \alpha_\agi \circ \mathsf{ob}_\agi$ in~$\G^\mathsf{K}$.

The following result establishes an important property of expanded games. 
Let $\G^{2\mathsf{K}}$ denote $(\G^\mathsf{K})^\mathsf{K}$.

\begin{lemma}
\label{lem:PDKforGKK}
Let~$\G$ be a \magii, $\G^\mathsf{K}$ be its MKBSC expansion, 
and $\G^{2\mathsf{K}}$ be the MKBSC expansion of~$\G^\mathsf{K}$.
Then, $\G^{2\mathsf{K}}$ fulfills the PDK condition. 
\end{lemma}

\begin{proof} 
As pointed out above, a sufficient condition on a game to guarantee 
that its MKBSC expansion fulfills the PDK condition is 
that no two distinct locations of the game are indistinguishable 
by all agents. This sufficient condition is enforced by the
partition step of Definition~\ref{def:mkbsc}, since two
knowledge states of the expansion can only be indistinguishable 
by all agents if they are equal. 

Formally, the proof proceeds by contradiction.
Assume that $\G^{2\mathsf{K}}$ does not fulfill the PDK condition.
By the definition of the PDK condition, there must then
be a knowledge state $\mathbf{s}$ of 
$\G^{2\mathsf{K}}$ such that 
$\cap_{\agi \in \Agt}\, \mathbf{s} (\agi)$ is not a singleton
set. And since the latter set, due to the pruning in the composition 
step of Definition~\ref{def:mkbsc}, also cannot be empty,
there must be (at least) two \emph{distinct} knowledge states 
$s_1$ and $s_2$ in $\G^\mathsf{K}$ 
such that $\set{s_1, s_2} \subseteq \mathbf{s} (\agi)$ for all 
$\agi \in \Agt$.
Hence, by the expansion step of Definition~\ref{def:mkbsc},
$s_1$ and $s_2$ must be indistinguishable in~$\G^\mathsf{K}$
for all $\agi \in \Agt$, and therefore, by the partition step of 
Definition~\ref{def:mkbsc},
$s_1 (\agi) = s_2 (\agi)$ for all $\agi \in \Agt$.
But then $s_1 = s_2$, and we arrive at a contradiction.
\end{proof}

This result will be important for the properties of the 
iterated construction studied in Section~\ref{sec:IteratedMKBSC}.

Strategy preservation in the reverse direction, from~$\G^\mathsf{K}$ to~$\G$, 
does \emph{not} depend on the PDK condition. 

\begin{theorem}[Reverse Strategy Preservation]
\label{thm:rev-strat-pres}
Let~$\G$ be a \magii, and let $\G^\mathsf{K}$ be its MKBSC expansion. 
Let~$\reach$ be an observable reachability objective in~$\G$, and 
$\reach^\mathsf{K}$ be its translation for~$\G^\mathsf{K}$. 
If there is a winning profile of observation-based perfect recall 
strategies in~$G^\mathsf{K}$ for~$\reach^\mathsf{K}$, then there 
is also  one in~$\G$ for~$\reach$. 
\end{theorem}

\begin{proof}
Informally, given a strategy profile~$\set{\alpha_\agi^\mathsf{K}}$ 
in~$\G^\mathsf{K}$, every agent~$\agi \in \Agt$ plays in~$\G$ by: 
\begin{enumerate}
\setlength{\itemsep}{-3pt}
\item 
recording the sequence of actions it has taken and observations 
it has made so far during the play, 
\item
following the unique path in~$\G^\mathsf{K}$ that corresponds 
to this sequence, and 
\item
for the corresponding sequence of observations in~$\G^\mathsf{K}$, 
taking the action as prescribed by~$\alpha_\agi^\mathsf{K}$ 
for that sequence.
\end{enumerate} 

Formally, let $\set{\alpha_\agi^\mathsf{K}}_{\agi \in \Agt}$, where 
$\alpha_\agi^\mathsf{K} : (\Obs^\mathsf{K}_\agi)^+ \rightarrow \Act_\agi$ 
for all~$\agi \in \Agt$, be a winning profile of observation-based 
perfect recall strategies in~$G^\mathsf{K}$ for the observable 
reachability objective~$\reach^\mathsf{K}$. 
For all $\agi \in \Agt$,
we define the functions $\alpha_\agi : \Obs_\agi^+ \rightarrow \Act_\agi$ 
by induction on the length of observation sequences. In the base case, define 
$\alpha_\agi (\set{\init}) \defeq \alpha_\agi^\mathsf{K} (\set{s_I})$. 
Let $\mu = o_0 o_1 \ldots o_m \in \Obs_\agi^+$ be an observation sequence, 
where $o_0 = \set{\init}$, and assume that $\alpha$~is defined for 
all its prefixes (induction hypothesis). Then, the actions 
$\sigma_k \defeq \alpha_\agi (\mu (k))$ are also defined for all 
$k : 0 \leq k \leq m$. Let $o_{m+1} \in \Obs_\agi$. 
By Lemma~\ref{lem:MKBSCCharacterisation}, the observation history 
$o_0 \sigma_0 o_1 \sigma_1 \ldots o_m \sigma_m o_{m+1}$ defines at most
one path 
$s_0 \sigma_0 s_1 \sigma_1 \ldots s_m \sigma_m s_{m+1}$ 
in~$\G^\mathsf{K}$. Define 
$\alpha_\agi (\mu \cdot o_{m+1}) \defeq 
 \alpha_i^\mathsf{K} (s_0 s_1 \ldots s_{m+1})$ 
if such a path exists (and otherwise its choice is immaterial). 
Thus the functions define a profile $\set{\alpha_\agi}_{\agi \in \Agt}$
of observation-based perfect recall strategies in~$\G$. 
We show that this profile is winning in~$\G$ for the 
objective~$\reach$.

Let $\pi = l_0 \sigma_0 l_1 \sigma_1 \ldots$ be an arbitrary outcome 
of~$\set{\alpha_\agi}_{\agi \in \Agt}$ in~$\G$.
Then, by Lemma~\ref{lem:MKBSCCharacterisation}, there is a play 
$\pi^\mathsf{K} = s_0 \sigma_0 s_1 \sigma_1 \ldots$ in~$G^\mathsf{K}$ 
such that $l_k \in \cap_{\agi \in \Agt}\, s_k (\agi)$ for all $k \geq 0$. 
By the definitions of~$\alpha_\agi$, 
$\pi^\mathsf{K}$ must be an outcome of~$\set{\alpha_i^\mathsf{K}}$ 
in~$\G^\mathsf{K}$, and must thus be winning for $\reach^\mathsf{K}$, i.e., 
there is an agent $\agi \in \Agt$, 
an index $k \geq 0$ and an observation 
$o \in \reach \cap \Obs_\agi$ such that $s_k (\agi) \subseteq o$.
But then there is also  an agent $\agi \in \Agt$
and an index $k \geq 0$ such that 
$\obsi{\agi}{l_k} \in \reach$, 
and hence $\pi$ is winning for~$\reach$. 
Since $\pi$ is arbitrary, the profile $\set{\alpha_\agi}_{\agi \in \Agt}$ 
must be winning in~$\G$ for the objective~$\reach$. 
\end{proof}

Again, it is easy to see from the proof that winning profiles of 
observation-based perfect recall strategies for observable 
safety objectives are also preserved, by replacing in the proof 
``there is an index $k \geq 0$'' with ``for all indices $k \geq 0$''. 

Further, note that the proof is constructive and reveals how observation-based 
memoryless strategies in~$\G^\mathsf{K}$ can be mapped to observation-based 
finite-memory strategies (i.e., transducers) in~$\G$ (which may, in some cases,
``degenerate'' to memoryless strategies in~$\G$). We will make use of 
this in Section~\ref{subsubsec:strategy-translation-transducers}.

\subsection{Strategy translation}
\label{subsec:strategy-translation}

While the results of the preceding subsection concern profiles of observation-based
perfect recall strategies, in this work we focus on searching for profiles 
of observation-based \emph{memoryless} strategies in the game~$\G^\mathsf{K}$.
For this class of strategies the synthesis problem has already been 
studied (see e.g.~\cite{DBLP:journals/logcom/PileckiBJ17}). 
If such a strategy profile can be found, it needs to be converted to an
observation-based strategy profile 
for play in the original game structure~$\G$. 
For this, we will offer here two solutions: 
\begin{itemize}
\item[($i$)]
the \emph{extensional solution}, where we convert each individual strategy 
to an individual observation-based finite-memory strategy (i.e., transducer), 
and
\item[($ii$)]
the \emph{intensional solution}, where we interpret the individual strategies as
knowledge-based strategies, based on a (common) knowledge 
representation and individual update functions. 
\end{itemize}
The two solutions will be shown to be equivalent, i.e., to give rise to the same 
sets of outcomes.

\subsubsection{Translation to transducers}
\label{subsubsec:strategy-translation-transducers}

We start with the important observation that, by virtue of how the
observations in~$\G^\mathsf{K}$ are defined in Definition~\ref{def:mkbsc},
every observation-based memoryless strategy for 
agent~$\agi$ in~$\G^\mathsf{K}$ is simultaneously a memoryless 
strategy in the game with perfect information~$(\G|_i)^\mathsf{K}$. 
This observation motivates the following construction,
which essentially combines each game~$(\G|_\agi)^\mathsf{K}$ and
individual memoryless strategy~$\strat^\mathsf{K}_\agi$ into a 
transducer~$A_\agi (\strat^\mathsf{K}_\agi)$
for agent~$\agi$ for play in~$\G$.

\begin{definition}[Induced Transducer]
\label{def:induced-transducer}
Let $\G = (\Agt, \Loc, \init, \Act, \Delta, \Obs)$ and for any 
$\agi \in \Agt$, let 
$(\G|_\agi)^\mathsf{K} = (S_\agi, s_{I,\agi}, \Act_\agi, \Delta^\mathsf{K}_\agi)$.
Let also~$\strat^\mathsf{K}_\agi : S_\agi \rightarrow \Act_\agi$ be a 
memoryless strategy in~$(\G|_\agi)^\mathsf{K}$. 
We define the following $\strat^\mathsf{K}_\agi$-\defstyle{induced transducer}:
\[ A_\agi (\strat^\mathsf{K}_\agi) \defeq 
(S_\agi, s_{I,\agi}, \Obs_\agi, \Act_\agi, \tau_\agi, \strat^\mathsf{K}_\agi) \]
where $\tau_\agi (s, o_\agi)$ is defined for $s \in S_\agi$ and $o_\agi \in \Obs_\agi$ as the unique $s' \in S_\agi$ such that $s' \subseteq o_\agi$ and $(s, \strat^\mathsf{K}_\agi (s), s') \in \Delta^\mathsf{K}_\agi$, if such an~$s'$
exists, and is undefined otherwise. 
\end{definition}
\noindent
Uniqueness of~$s'$ in the definition is guaranteed by 
Lemma~\ref{lem:MKBSCCharacterisation}. 

The transducer $A_\agi (\strat^\mathsf{K}_\agi)$~is, by 
Definition~\ref{def:finite-memory-strategy}, 
an observation-based finite-memory strategy for agent~$\agi$ in~$\G$.
The transducer can be \emph{pruned} by removing, 
from each memory state~$s$, the outgoing edges for actions other 
than~$\alpha^\mathsf{K}_\agi (s)$, then by removing the unreachable 
memory states, and finally by abstracting away the structure of~$s$ 
(since only the identity of the memory states is relevant). 

\begin{theorem}[Strategy Correspondence]
\label{thm:strategy-correspondence}
Let~$\G$  
be a \magii for a set of agents~$\Agt$, and let   
$\G^\mathsf{K}$ 
be its MKBSC expansion. 
Let~$\reach$ be an observable reachability objective in~$\G$, and $\reach^\mathsf{K}$  
be its translation in~$\G^\mathsf{K}$. 
Finally, let $\set{\alpha^\mathsf{K}_\agi}_{\agi \in \Agt}$ be a profile 
of observation-based memoryless strategies in~$\G^\mathsf{K}$, 
and~$\set{A_\agi (\strat^\mathsf{K}_\agi)}_{\agi \in \Agt}$ be the 
corresponding profile of induced transducers for~$\G$.
\begin{itemize}
\item[($i$)]
If $\set{A_\agi (\strat^\mathsf{K}_\agi)}_{\agi \in \Agt}$ 
is winning for~$\reach$ in~$\G$,
and $\G^\mathsf{K}$ fulfills the PDK condition, 
then $\set{\alpha^\mathsf{K}_\agi}_{\agi \in \Agt}$ is winning 
for~$\reach^\mathsf{K}$ in~$\G^\mathsf{K}$.
\item[($ii$)]
If $\set{\alpha^\mathsf{K}_\agi}_{\agi \in \Agt}$ is winning for~$\reach^\mathsf{K}$ 
in~$\G^\mathsf{K}$, then $\set{A_\agi (\strat^\mathsf{K}_\agi)}_{\agi \in \Agt}$ 
is winning for~$\reach$ in~$\G$.
\end{itemize}
\end{theorem}

\begin{proof}
($i$)
The proof adapts the strategy construction used in the proof of 
Theorem~\ref{thm:strat-pres}. 
Let $\set{A_\agi (\strat^\mathsf{K}_\agi)}_{\agi \in \Agt}$ 
be winning for~$\reach$ in~$\G$, and let $\G^\mathsf{K}$ fulfill
the PDK condition.
Let $\pi^\mathsf{K} = s_0 \sigma_0 s_1 \sigma_1 \ldots$ be an arbitrary outcome 
of $\set{\alpha_\agi^\mathsf{K}}_{\agi \in \Agt}$ in~$\G^\mathsf{K}$.
Then, by Lemma~\ref{lem:MKBSCCharacterisation} and since $G^\mathsf{K}$ 
fulfills the PDK condition, the sequence
$\pi = l_0 \sigma_0 l_1 \sigma_1 \ldots$ such that 
$\set{l_k} = \cap_{\agi \in \Agt}\, s_k (\agi)$ for all $k \geq 0$, 
must be a full play  in~$\G$. 
Now, by Definition~\ref{def:induced-transducer}, 
$\pi$ must be an outcome of
$\set{A_\agi (\strat^\mathsf{K}_\agi)}_{\agi \in \Agt}$ in~$\G$. 
Since $\set{A_\agi (\strat^\mathsf{K}_\agi)}_{\agi \in \Agt}$ 
is winning for~$\reach$ in~$\G$,
$\pi$ must be winning for~$\reach$ in~$\G$,
and hence, by the definition of~$\reach^\mathsf{K}$, 
$\pi^\mathsf{K}$ must be winning for~$\reach^\mathsf{K}$ in~$\G^\mathsf{K}$.
But~$\pi^\mathsf{K}$ is arbitrary, and therefore 
$\set{\alpha^\mathsf{K}_\agi}_{\agi \in \Agt}$ must be winning 
for~$\reach^\mathsf{K}$ in~$\G^\mathsf{K}$. 

($ii$)
The proof adapts the strategy construction used in the proof of 
Theorem~\ref{thm:rev-strat-pres}, using the observation that 
observation-based memoryless strategies for agent~$\agi$ 
in~$\G^\mathsf{K}$ correspond to memoryless strategies 
in~$(\G|_i)^\mathsf{K}$.

Let $\set{\alpha^\mathsf{K}_\agi}_{\agi \in \Agt}$ be winning for~$\reach^\mathsf{K}$ 
in~$\G^\mathsf{K}$. Let $\pi = l_0 l_1 l_2 \ldots$ be an arbitrary outcome 
of $\set{A_\agi (\strat^\mathsf{K}_\agi)}_{\agi \in \Agt}$ in~$\G$. This
outcome induces a corresponding sequence of joint observations, from which,
using $\set{A_\agi (\strat^\mathsf{K}_\agi)}_{\agi \in \Agt}$, one 
can recover the corresponding sequence of joint actions. By 
Lemma~\ref{lem:MKBSCCharacterisation}, these two sequences (of 
joint actions and joint observations) give rise to a unique 
play~$\pi^\mathsf{K}$ in~$\G^\mathsf{K}$, the individual knowledge 
states of which are subsets of the corresponding individual 
observations. 
Now, by Definition~\ref{def:induced-transducer}, 
$\pi^\mathsf{K}$ must be an outcome of
$\set{\alpha^\mathsf{K}_\agi}_{\agi \in \Agt}$ in~$\G^\mathsf{K}$. 
Since $\set{\alpha^\mathsf{K}_\agi}_{\agi \in \Agt}$ is winning 
for~$\reach^\mathsf{K}$ in~$\G^\mathsf{K}$, $\pi^\mathsf{K}$ must be 
winning for~$\reach^\mathsf{K}$ in~$\G^\mathsf{K}$,
and hence, by the definition of~$\reach^\mathsf{K}$, $\pi$ must 
be winning for~$\reach$ in~$\G$. But~$\pi$ is arbitrary, and
therefore $\set{A_\agi (\strat^\mathsf{K}_\agi)}_{\agi \in \Agt}$ 
must be winning for~$\reach$ in~$\G$. 
\end{proof}

The above results suggest a \emph{method to synthesise observation-based
finite-memory strategies} for reachability objectives~$\reach$ in a 
\magii~$\G$, based on: 

\begin{itemize}
\item[($\mathit{i}$)]~computing the MKBSC 
expansion~$\G^\mathsf{K}$ of the game, 

\item[($\mathit{ii}$)]~searching 
for a winning profile of observation-based memoryless strategies 
(for the translated objective~$\reach^\mathsf{K}$) there, and if such 
a profile $\set{\alpha^\mathsf{K}_\agi}_{\agi \in \Agt}$ is found,

\item[($\mathit{iii}$)]~translating the latter back in the form of the 
transducers $\set{A_\agi (\strat^\mathsf{K}_\agi)}_{\agi \in \Agt}$.
\end{itemize}

\subsubsection{Translation to knowledge-based strategies}
\label{subsubsec:strategy-translation-knowledge-based strategies}

To be able to interpret the individual strategies~$\strat^\mathsf{K}_\agi$ 
of the agents in~$\G^\mathsf{K}$ as individual knowledge-based strategies 
in~$\G$, following the framework outlined in Section~\ref{subsec:kb-strat},
we need to define a knowledge representation and individual 
knowledge update functions. 
As a \defstyle{first-order knowledge representation} structure we 
will use non-empty sets $A \subseteq \Loc$ 
of locations in~$\G$. For a given agent, 
the intended interpretation of such a set is as 
the agent's \emph{most precise estimate of the actual location 
of the game} and, thus, represents the agent's exact uncertainty about the
actual state-of-affairs.
Given this knowledge representation, each memoryless strategy
$\strat^\mathsf{K}_\agi : S_\agi \rightarrow \Act_\agi$
in~$(\G|_i)^\mathsf{K}$ can simultaneously be viewed as an 
individual \defstyle{first-order knowledge-based 
strategy} for agent~$\agi$ in~$\G$. 

\begin{definition}[Knowledge Update]
\label{def:knowledge-update}
For $s \in 2^{\Loc} \setminus \set{\varnothing}$, 
$\act_\agi \in \Act_\agi$ and  $o_\agi \in \Obs_\agi$, 
the \defstyle{knowledge update} 
function of agent~$\agi \in \Agt$ is defined as follows: 
$$ \delta_\agi (s, \act_\agi, o_\agi) \defeq
   \setdef{l' \in o_\agi}{\exists \act \in \Act.\ 
      (\act (\agi) = \act_\agi \wedge \exists l \in s.\ 
         ((l, \act, l') \in \Delta))} $$
if the set is non-empty, and is undefined otherwise. 
\end{definition}

\noindent
The set is the most precise estimate of agent~$\agi$ of the new actual 
location upon taking
the action~$\sigma_\agi$ and making the new observation~$o_\agi$. 
Note that the update functions~$\delta_\agi$ do \emph{not} depend 
on~$\strat^\mathsf{K}_\agi$ (which may not be the case in the
\textbf{(YN)} and \textbf{(YY)} cases discussed in Section~\ref{subsec:KnowledgeBasedStrategies}).
The \defstyle{initial knowledge} of each agent~$\agi$ is~$\set{\init}$.

As the following result states, the first-order knowledge-based 
strategies defined in this way agree with the finite-memory ones from 
Definition~\ref{def:induced-transducer}.

\begin{theorem}[Strategy Equivalence]
\label{thm:strategy-equivalence}
Let $\G$  
be a \magii,  
$\G^\mathsf{K}$ 
be its MKBSC expansion,
$\set{\alpha^\mathsf{K}_\agi}_{\agi \in \Agt}$ be a profile 
of observation-based memoryless strategies in~$\G^\mathsf{K}$, and
$\set{A_\agi (\strat^\mathsf{K}_\agi)}_{\agi \in \Agt}$ be the 
corresponding profile of induced transducers for~$\G$. 
Then, the strategy profile 
$\set{A_\agi (\strat^\mathsf{K}_\agi)}_{\agi \in \Agt}$,
and the profile of first-order knowledge-based strategies based on 
$\set{\alpha^\mathsf{K}_\agi}_{\agi \in \Agt}$ and 
$\set{\delta_\agi}_{\agi \in \Agt}$, give rise to the same
set of outcomes in~$\G$.
\end{theorem}

\begin{proof}
We show that
$\tau_\agi (s, o_\agi) = \delta_\agi (s, \alpha^\mathsf{K}_\agi (s), o_\agi)$
for all $s \in 2^{\Loc} \setminus \set{\varnothing}$ 
and $o_\agi \in \Obs_\agi$.
The result then follows from Definition~\ref{def:induced-transducer},
Definition~\ref{def:finite-memory-strategy}, and the
definition of first-order knowledge-based strategies.

Let $s \in 2^{\Loc} \setminus \set{\varnothing}$ 
and $o_\agi \in \Obs_\agi$. We have:
$$ \begin{array}{cll}
   & \tau_\agi (s, o_\agi) & \\
   = & 
     \mbox{the unique $s' \in S_\agi$ such that $s' \subseteq o_\agi$ and 
     $(s, \strat^\mathsf{K}_\agi (s), s') \in \Delta^\mathsf{K}_\agi$} &
     \{\mbox{Def.~\ref{def:induced-transducer}}\} \\
   = &
     \setdef{l' \in o_\agi}
        {\exists l \in s.\ (l, \strat^\mathsf{K}_\agi (s), l') \in \Delta_\agi} &
     \{\mbox{Def.~\ref{def:mkbsc}.2}\} \\
   = &
     \setdef{l' \in o_\agi}
        {\exists \act \in \Act.\ (\act (\agi) = \strat^\mathsf{K}_\agi (s) 
         \wedge \exists l \in s.\ (l, \act, l') \in \Delta)} &
     \{\mbox{Def.~\ref{def:mkbsc}.1}\} \\
   = &
     \delta_\agi (s, \alpha^\mathsf{K}_\agi (s), o_\agi) &
     \{\mbox{Def.~\ref{def:knowledge-update}}\} \\
   \end{array} $$
if such an~$s'$ exists; otherwise, by 
Lemma~\ref{lem:MKBSCCharacterisation}, also
$\delta_\agi (s, \alpha^\mathsf{K}_\agi (s), o_\agi)$
is undefined.
\end{proof}

As a corollary of Theorems~\ref{thm:strategy-correspondence} 
and~\ref{thm:strategy-equivalence}, we have that:
\begin{itemize}
\item[($i$)]
if $\set{\alpha^\mathsf{K}_\agi}_{\agi \in \Agt}$ 
with $\set{\delta_\agi}_{\agi \in \Agt}$ 
is winning for~$\reach$ in~$\G$,
and $\G^\mathsf{K}$ fulfills the PDK condition, 
then $\set{\alpha^\mathsf{K}_\agi}_{\agi \in \Agt}$ is winning 
for~$\reach^\mathsf{K}$ in~$\G^\mathsf{K}$, and
\item[($ii$)]
if $\set{\alpha^\mathsf{K}_\agi}_{\agi \in \Agt}$ is winning for~$\reach^\mathsf{K}$ 
in~$\G^\mathsf{K}$, then $\set{\alpha^\mathsf{K}_\agi}_{\agi \in \Agt}$ 
with $\set{\delta_\agi}_{\agi \in \Agt}$ 
is winning for~$\reach$ in~$\G$.
\end{itemize}

Every profile of first-order knowledge-based strategies 
in~$\G$ is at the same time a profile of observation-based memoryless 
strategies in~$\G^\mathsf{K}$, and vice versa.
Then, for a given \magii~$\G$ and observable reachability 
objective~$\reach$, if $\G^\mathsf{K}$ fulfills the PDK condition,
a winning profile of first-order knowledge-based 
strategies exists if and only if a winning profile of observation-based 
memoryless strategies exists in~$\G^\mathsf{K}$ for~$\reach^\mathsf{K}$.

Our strategy synthesis method is, therefore, \emph{complete} under PDK for the class of first-order knowledge-based strategies with respect to observable reachability objectives, in the sense that if a winning profile of first-order knowledge-based strategies exists, it will be found with our method.

\begin{example}
\label{ex:kb-strat}
Consider again our running Example~\ref{ex:robots}, and note that the
PDK condition holds for the MKBSC expansion of this game. 
Let the joint observable objective (in~$\G$) be to reach location $\mathsf{good}$
(and recall that it suffices for just one of the robots to observe this).
Here is a winning profile of first-order knowledge-based strategies for
play in~$\G$, 
where robot~0 follows the strategy defined in the left two columns
of the table below, and updates its knowledge according to~$\delta_0$, 
partially shown in the right two columns:

$$ \begin{array}{|c||c||c|c|} 
   \hline
   \mathit{Knowledge~state} & \mathit{Action} & 
   \mathit{On~observing}~\set{\mathsf{bad}} & 
   \mathit{On~observing}~\set{\mathsf{good}} \\
   \hline\hline

   \set{\mathsf{start}} & 
   \mathsf{grab} & 
   \set{\mathsf{bad}} & 
   \set{\mathsf{good}} \\
   \hline

   \set{\mathsf{bad}} & 
   \mathsf{squeeze} & 
   \mathsf{NA} & 
   \set{\mathsf{good}} \\
   \hline

   \set{\mathsf{good}} & 
   \mathsf{squeeze} & 
   \mathsf{NA} & 
   \set{\mathsf{good}} \\
   \hline

   \end{array} $$~

\noindent
while robot~1 follows the strategy defined in the left two columns 
of the table below, and updates its knowledge according to~$\delta_1$, 
partially shown in the right column:

$$ \begin{array}{|c||c||c|} 
   \hline
   \mathit{Knowledge~state} & \mathit{Action} & 
   \mathit{On~observing}~\set{\mathsf{good}, \mathsf{bad}} \\
   \hline\hline

   \set{\mathsf{start}} & 
   \mathsf{grab} & 
   \set{\mathsf{good}, \mathsf{bad}}  \\
   \hline

   \set{\mathsf{good}, \mathsf{bad}} & 
   \mathsf{squeeze} & 
   \set{\mathsf{good}}  \\
   \hline

   \end{array} $$~

\noindent
If, however, the objective is to reach location $\mathsf{win}$, then there 
is \emph{no winning profile of first-order knowledge-based strategies}. 
Intuitively, the reason for this is that robot~0 is obliged to take the 
same action in both locations of~$\G^\mathsf{K}$ where it knows 
$\set{\mathsf{good}}$, but $\mathsf{win}$ can only be reached by taking 
different actions.
This problem will be resolved below with the help of second-order 
knowledge-based strategies. 
\end{example}

The \emph{duality} between the intensional and the extensional views 
exhibited above is not surprising. 
Having a knowledge-based strategy in the form of 
$\alpha^\mathsf{K}_\agi$ and $\delta_\agi$, agent~$\agi$ can
reconstruct the transducer~$A_\agi (\alpha^\mathsf{K}_\agi)$, as evidenced 
by the proof of Theorem~\ref{thm:strategy-equivalence}.
One can thus view the execution of an individual knowledge-based strategy 
by an agent as constructing the corresponding pruned
transducer  \emph{on-the-fly}.

\subsection{Strategy synthesis}
\label{subsec:strategy-synthesis}

In the beginning of Section~\ref{subsubsec:strategy-translation-transducers}
we noted that every observation-based memoryless strategy 
$\alpha^\mathsf{K}_\agi$ for agent~$\agi$ in~$\G^\mathsf{K}$ is also 
a memoryless strategy in the single-agent game with perfect information~$(\G|_i)^\mathsf{K}$.
This fact can also be useful for the \emph{synthesis} of profiles 
of observation-based memoryless strategies in~$\G^\mathsf{K}$, since the
synthesis of memoryless strategies in single-agent games of perfect information is well-studied.
For instance, in the context of reachability objectives the standard 
synthesis technique is based on the notion of \emph{controllable predecessors}
(see e.g.~\cite{DBLP:books/cu/11/0001R11}).

Another useful fact is that 
if a profile $\set{\alpha^\mathsf{K}_\agi}_{\agi \in \Agt}$
of observation-based memoryless strategies in~$\G^\mathsf{K}$ is winning
for a (translated) reachability objective~$\reach^\mathsf{K}$, then,
due to the consistency of the joint knowledge states of~$\G^\mathsf{K}$,
every individual memoryless strategy~$\alpha^\mathsf{K}_\agi$ has an outcome 
$\pi_i = s_0 s_1 s_2 \ldots$ in~$(\G|_i)^\mathsf{K}$ such that 
$\exists r \geq 0.\ \exists o \in \reach.\ s_r \cap o \neq \varnothing$.
This suggests the following \textit{simple heuristic} for synthesis of 
profiles of observation-based memoryless strategies in~$\G^\mathsf{K}$: 
\begin{enumerate}
\item
For each agent $\agi \in \Agt$, find a memoryless strategy 
$\alpha^\mathsf{K}_\agi$ in~$(\G|_i)^\mathsf{K}$ that has a
winning outcome for the reachability objective 
$\reach_\agi^\mathsf{K} = \setdef{s \in S_\agi}{\exists o \in \reach.\ s \cap o \neq \varnothing}$. 
\item
Check whether the profile $\set{\alpha^\mathsf{K}_\agi}_{\agi \in \Agt}$ is
winning for~$\reach^\mathsf{K}$. If it is not, backtrack to step~1.
\end{enumerate}

To implement the first step of the heuristic, one can adapt the notion 
of controllable predecessors to finding strategies where \emph{some} outcome 
is winning (rather than \emph{all} outcomes, as the standard formulation achieves).
For the second step, it suffices to simulate the profile 
$\set{\alpha^\mathsf{K}_\agi}_{\agi \in \Agt}$ on the game~$\G^\mathsf{K}$,
following each outcome up to the first knowledge state which either 
belongs to~$\reach^\mathsf{K}$, or else has already been visited by 
the outcome. In the latter case, the profile is not winning. Since
the set of knowledge states is finite, this check terminates.

\section{The iterated MKBSC construction} 
\label{sec:IteratedMKBSC}

Applying the MKBSC to a \magii $\G$ does not necessarily result in a game 
with perfect information, but in general produces another \magii. Thus, 
as it was first observed in~\cite{lun-17-msc} by the third co-author, 
the construction can always be applied again, 
iteratively, producing an infinite  (but possibly collapsing) hierarchy of expansions $\G^{\mathsf{K}}$, $\G^{2\mathsf{K}}$, $\G^{3\mathsf{K}}$, \ldots. 
We show here that such repeated application produces a hierarchy of higher-order knowledge representation structures for the agents in the team, and using these can increase its strategic abilities.  

\begin{figure}[ht]
    \centering
    \includegraphics[scale=.28]{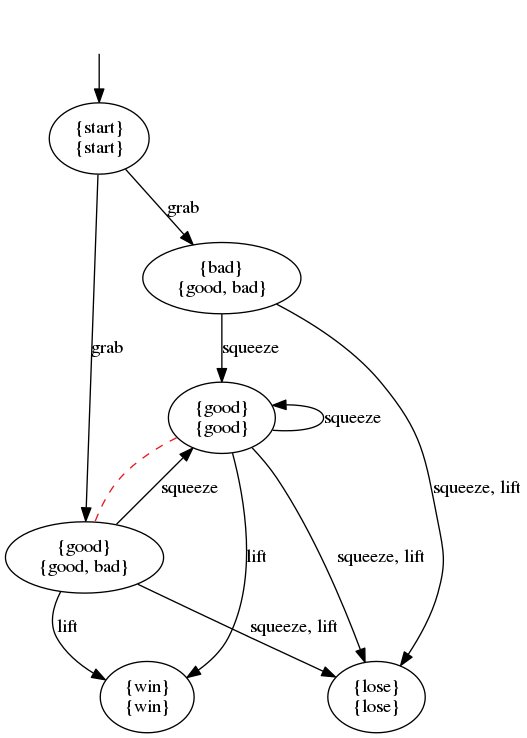}
    \hspace{5truemm}
    \includegraphics[scale=.28]{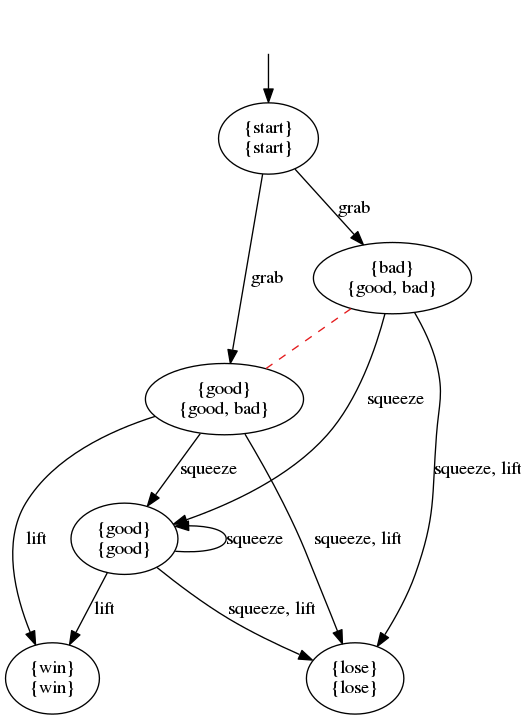}
    \caption{The individual projections $\G^\mathsf{K}|_0$ and~$\G^\mathsf{K}|_1$.}
    \label{fig:rep-indiv-proj}
\end{figure}

\begin{figure}[ht]
    \centering
    \includegraphics[scale=.28]{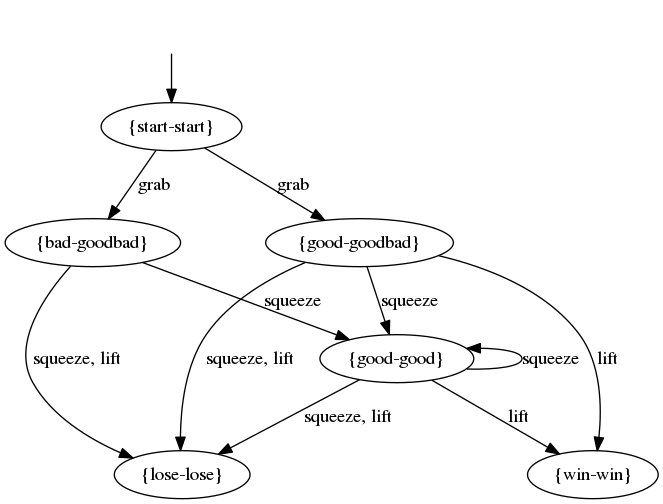}
    \hspace{5truemm}
    \includegraphics[scale=.28]{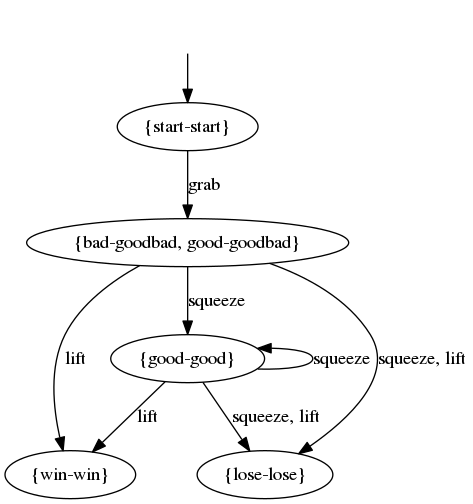}
    \caption{The individual expansions $(\G^\mathsf{K}|_0)^\mathsf{K}$ 
             and~$(\G^\mathsf{K}|_1)^\mathsf{K}$.}
    \label{fig:rep-indiv-exp}
\end{figure}

\begin{figure}[ht]
    \centering
    \includegraphics[scale=.35]{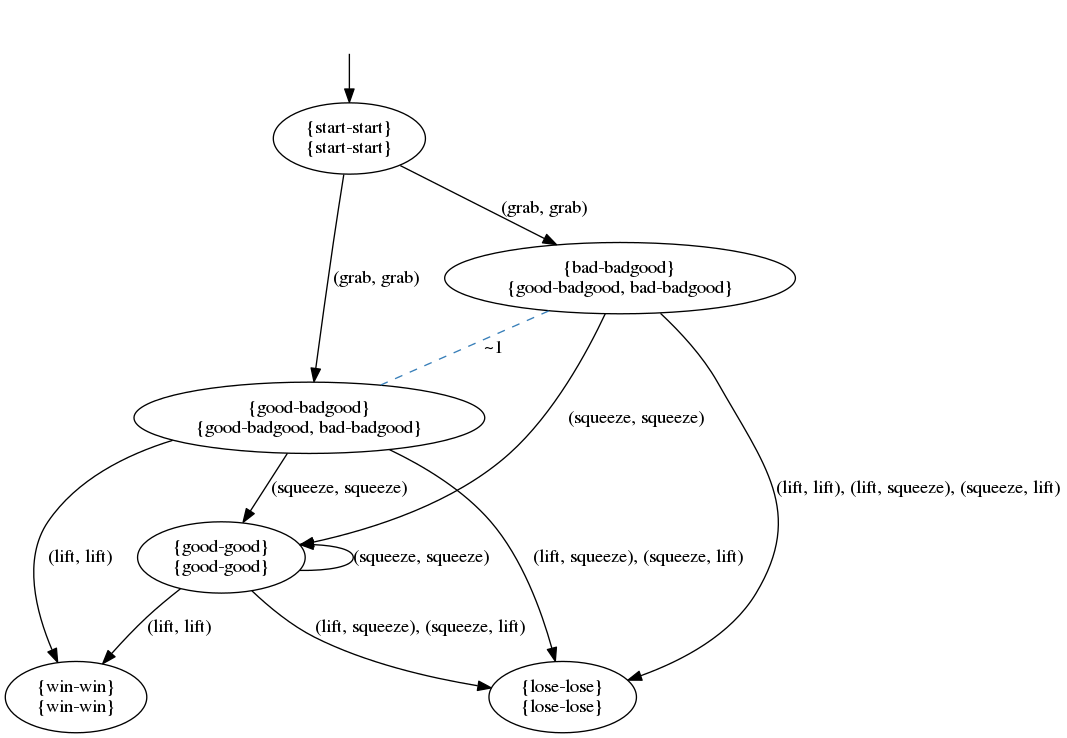}
    \caption{The pruned product $\G^{2\mathsf{K}}$.}
    \label{fig:rep-prod}
\end{figure}

\begin{example}
\label{ex:iter-kb-strat} 
We apply below the MKBSC construction on the game~$\G^\mathsf{K}$ 
from Figure~\ref{fig:prod} to produce~$\G^{2\mathsf{K}}$. 
The individual projections $\G^\mathsf{K}|_0$ and~$\G^\mathsf{K}|_1$ are
shown in Figure~\ref{fig:rep-indiv-proj}, with corresponding expansions 
$(\G^\mathsf{K}|_0)^\mathsf{K}$ and~$(\G^\mathsf{K}|_1)^\mathsf{K}$,
as shown in Figure~\ref{fig:rep-indiv-exp}.
The pruned product~$\G^{2\mathsf{K}}$ is shown in 
Figure~\ref{fig:rep-prod}.

Now, for the robot team to reach location $\mathsf{win}$ in~$\G$,  
robot~0 can follow the second-order knowledge-based strategy (extracted
from the above games as described below): 

$$ \begin{array}{|c||c|} 
   \hline
   \mathit{Knowledge~state} & \mathit{Action} \\
   \hline\hline

   \set{(\set{\mathsf{start}}, \set{\mathsf{start}})} & \mathsf{grab} \\
   \hline

   \set{(\set{\mathsf{good}}, \set{\mathsf{bad}, \mathsf{good}})} & \mathsf{squeeze} \\
   \hline

   \set{(\set{\mathsf{bad}}, \set{\mathsf{bad}, \mathsf{good}})} & \mathsf{squeeze} \\
   \hline

   \set{(\set{\mathsf{good}}, \set{\mathsf{good}})} & \mathsf{lift} \\
   \hline

   \end{array} $$~

\noindent
while the strategy of robot~1 follows the table:
$$ \begin{array}{|c||c|} 
   \hline
   \mathit{Knowledge~state} & \mathit{Action} \\
   \hline\hline

   \set{(\set{\mathsf{start}}, \set{\mathsf{start}})} & \mathsf{grab} \\
   \hline

   \set{(\set{\mathsf{bad}}, \set{\mathsf{bad}, \mathsf{good}}), (\set{\mathsf{good}}, \set{\mathsf{bad}, \mathsf{good}})} & \mathsf{squeeze} \\
   \hline

   \set{(\set{\mathsf{good}}, \set{\mathsf{good}})} & \mathsf{lift} \\
   \hline

   \end{array} $$~

\noindent
This means, for instance, that robot~0 will squeeze whenever 
it \emph{knows} that robot~1 (the one without a grip sensor) 
is uncertain about whether the grip is good or not. 
\end{example}

\subsection{Generalised induced transducers}
\label{subsec:generalised-induced-transducers}

Now, we discuss the general case of \emph{iterating} the MKBSC 
construction $j$~times, for any $j\geq 1$, resulting in the expanded 
game structure~$\G^{j\mathsf{K}}$, and generalise our results from 
the preceding section. 
Let~$\G^{0\mathsf{K}}$ denote the original game~$\G$.

Intuitively, the single-agent game with perfect information
$(\G^{j\mathsf{K}}|_\agi)^\mathsf{K}$ represents the possible
``dynamics'' of agent~$\agi$'s $(j+1)$-order knowledge. 
Similarly to the construction presented in 
Definition~\ref{def:induced-transducer}, one can combine 
each~$(\G^{j\mathsf{K}}|_\agi)^\mathsf{K}$ and
individual memoryless strategy~$\strat^{(j+1)\mathsf{K}}_\agi$ 
into a transducer~$A_\agi (\strat^{(j+1)\mathsf{K}}_\agi)$
for agent~$\agi$ for play in~$\G$. 
To achieve this, however, we first need to formally connect,
for every agent~$\agi \in \Agt$, the  knowledge states 
of~$(\G^{j\mathsf{K}}|_\agi)^\mathsf{K}$ to the 
observations~$\Obs_\agi$ of that agent in~$\G$. 
This can be achieved by observing that each knowledge state~$s_\agi \in S_\agi$ 
of~$(\G^{j\mathsf{K}}|_\agi)^\mathsf{K}$ is a set of locations 
in~$\G^{j\mathsf{K}}$, which, when $j>0$, are tuples
that agree on their $\agi$-th component,
where this $\agi$-th component is in turn a knowledge state
in~$(\G^{(j-1)\mathsf{K}}|_\agi)^\mathsf{K}$. 
We can thus repeat this process until reaching a set of locations in~$\G$. 
Let us denote this set by~$\hat{s_\agi} \subseteq \Loc$. 
\label{def:hat-s}
By virtue of the MKBSC construction, $\hat{s_\agi}$~will be non-empty, 
and will be a subset of some observation $o_\agi \in \Obs_\agi$ in the 
original game~$\G$.  

For the expanded game structures~$\G^{j\mathsf{K}}$,
the connection between the joint knowledge states~$s$ in the latter
and the locations in~$\G$ is established via iterated intersection
of the sets comprising the tuples in~$s$ until obtaining a set of
locations in~$\G$. Let us denote this set by~$\Cap s \subseteq \Loc$.
Iterated intersection is well-defined by virtue of 
Lemma~\ref{lem:PDKforGKK}.

The following result generalises Lemma~\ref{lem:MKBSCCharacterisation}.

\begin{lemma}
\label{lem:iMKBSCCharacterisation}
Let $\G = (\Agt, \Loc, \init, \Act, \Delta, \Obs)$ be a \magii for a 
set of agents~$\Agt$, let $j \geq 1$, and let 
$\G^{j\mathsf{K}} = 
 (\Agt, S^{j\mathsf{K}}, s_I, \Act, \Delta^{j\mathsf{K}}, \Obs^{j\mathsf{K}})$ 
be the $j$-iterated MKBSC expansion of~$\G$. 
Further, let $s \in S^{j\mathsf{K}}$, $\act \in \Act$ and $o \in \Obs^p$.
Define the set:
$$ X^{(j)} \>\defeq\> 
     \setdef{l' \in \cap_{\agi \in \Agt}\,  o (\agi)}
            {\exists l \in \Cap s.\ (l, \act, l') \in \Delta} $$
Then, $X^{(j)}$ is non-empty if and only if there is 
$s' \in S^{j\mathsf{K}}$ such that 
$(s, \act, s') \in \Delta^{j\mathsf{K}}$ 
and such that for all $\agi \in \Agt$, 
$\hat{s}' (\agi) \subseteq o (\agi)$.
This $s' \in S$ is then unique and we have
$X^{(j)} \subseteq \Cap s'$,
and if $\G^\mathsf{K}$ fulfills the PDK condition, 
we have $X^{(j)} = \Cap s'$.
\end{lemma}

\begin{proof}
By mathematical induction on~$j$.
The base case of~$j=1$ is established by 
Lemma~\ref{lem:MKBSCCharacterisation}.
Assume (as the induction hypothesis) that the result holds for~$j$.
We show that the result then follows for~$j+1$.

Let $s \in S^{(j+1)\mathsf{K}}$, $\act \in \Act$ and $o \in \Obs^p$.
Further, let $X^{(j+1)}$ be defined as above.
Consider the case when $X^{(j+1)}$ is non-empty. 
(The case when $X^{(j+1)}$ is empty is analogous.)
Then, there must be locations $l, l' \in \Loc$ in~$\G$ such that
$l \in \Cap s$, $l' \in \cap_{\agi \in \Agt}\,  o (\agi)$ and
$(l, \act, l') \in \Delta$. 
Let $s_1$ denote the sole element of the (simple) intersection of 
the sets comprising the tuple~$s$ (by Lemma~\ref{lem:PDKforGKK}, 
this intersection must be a sigleton set). Then 
$s_1 \in S^{j\mathsf{K}}$ and $l \in \Cap s_1$ must be the case. 
Let $X^{(j)}$ be defined as above (but w.r.t.\ $s_1 \in S^{j\mathsf{K}}$).
By the induction hypothesis, $X^{(j)}$ must also be non-empty,
and hence, there is exactly one $s_1' \in S^{j\mathsf{K}}$ such that 
$(s_1, \act, s_1') \in \Delta^{j\mathsf{K}}$ 
and for all $\agi \in \Agt$, $\hat{s_1}' (\agi) \subseteq o (\agi)$.
Then, $l' \in \Cap s_1'$ must be the case.

Now, let $o_1 \in \Obs^{j\mathsf{K}}$ be the unique observation 
profile in~$\G^{j\mathsf{K}}$ that the agents make in~$s_1'$. 
By Lemma~\ref{lem:MKBSCCharacterisation}, 
there is exactly one $s' \in S^{(j+1)\mathsf{K}}$ such that 
$(s, \act, s') \in \Delta^{(j+1)\mathsf{K}}$ 
and for all $\agi \in \Agt$, $s' (\agi) \subseteq o_1 (\agi)$.
Then, for all $\agi \in \Agt$, $\hat{s}' (\agi) \subseteq o (\agi)$
must be the case.
Furthermore, by Lemma~\ref{lem:MKBSCCharacterisation} and 
Lemma~\ref{lem:PDKforGKK}, $s_1'$ must be the sole element of 
the intersection of the sets comprising the tuple~$s'$, and
thus, since $l' \in \Cap s_1'$, also $l' \in \Cap s'$ must
be the case. This establishes $X^{(j+1)} \subseteq \Cap s'$.

Finally, if $\G^\mathsf{K}$ fulfills the PDK condition,
because of Lemma~\ref{lem:PDKforGKK} all iterated intersections
used in the proof must be singletons, and thus $X^{(j+1)} = \Cap s'$.
\end{proof}
\noindent
As before, this result lifts naturally to observation histories:
every sequence~$\pi$ of joint actions and joint observations of~$\G$ 
(where~$\pi$ is not necessarily a path in~$\G$) gives rise to at most 
one path in~$\G^{j\mathsf{K}}$ such that at each corresponding step 
of the two sequences, $\hat{s} (\agi) \subseteq o(\agi)$ holds for 
all $\agi \in \Agt$.
Furthermore, every full play~$\pi$ in~$\G$ gives rise to exactly one 
full play in~$\G^{j\mathsf{K}}$ that is consistent with the actions 
and the observations of the agents. 

\begin{definition}[Generalised Induced Transducer]
\label{def:generalised-induced-transducer}
Let $\G = (\Agt, \Loc, \init, \Act, \Delta, \Obs)$ and for any 
$\agi \in \Agt$ and $j \geq 0$, let 
$(\G^{j\mathsf{K}}|_\agi)^\mathsf{K} = (S_\agi, s_{I,\agi}, \Act_\agi, \Delta^{(j+1)\mathsf{K}}_\agi)$.
Let also~$\strat^{(j+1)\mathsf{K}}_\agi : S_\agi \rightarrow \Act_\agi$ be a 
memoryless strategy in~$(\G^{j\mathsf{K}}|_\agi)^\mathsf{K}$. 
We define the $\strat^{(j+1)\mathsf{K}}_\agi$-\defstyle{induced transducer}
as:
\[ A_\agi (\strat^{(j+1)\mathsf{K}}_\agi) \defeq 
(S_\agi, s_{I,\agi}, \Obs_\agi, \Act_\agi, \tau^{j+1}_\agi, \strat^{(j+1)\mathsf{K}}_\agi) \]
where $\tau^{j+1}_\agi (s, o_\agi)$ is defined for $s \in S_\agi$ and 
$o_\agi \in \Obs_\agi$ as the unique
$s' \in S_\agi$ such that $\hat{s}' \subseteq o_\agi$ and 
$(s, \strat^{(j+1)\mathsf{K}}_\agi (s), s') \in \Delta^{(j+1)\mathsf{K}}_\agi$,
if such an~$s'$ exists, and is undefined otherwise. 
\end{definition}

The transducer $A_\agi (\strat^{(j+1)\mathsf{K}}_\agi)$~is, by 
Definition~\ref{def:finite-memory-strategy}, 
an observation-based finite-memory strategy for agent~$\agi$ in~$\G$.
The transducer can be \emph{pruned} by removing, 
from each memory state~$s$, the outgoing edges for actions other 
than~$\strat^{(j+1)\mathsf{K}}_\agi (s)$, then by removing the unreachable 
memory states, and finally by abstracting away the structure of~$s$ 
(since only the identity of the memory states is relevant). 

The following result generalises Theorem~\ref{thm:strategy-correspondence}
to the iterated MKBSC.

\begin{theorem}[Generalised Strategy Correspondence]
\label{thm:gen-strategy-correspondence}
Let~$\G$ be a \magii for a set of agents~$\Agt$, let $j \geq 1$, 
and let $\G^{j\mathsf{K}}$ be the $j$-iterated MKBSC expansion of~$\G$. 
Let~$\reach$ be an observable reachability objective in~$\G$, and 
$\reach^{j\mathsf{K}}$ be its $j$-iterated translation in~$\G^{j\mathsf{K}}$. 
Finally, let $\set{\alpha^{j\mathsf{K}}_\agi}_{\agi \in \Agt}$ be a profile 
of observation-based memoryless strategies in~$\G^{j\mathsf{K}}$, 
and~$\set{A_\agi (\strat^{j\mathsf{K}}_\agi)}_{\agi \in \Agt}$ be the 
corresponding profile of generalised induced transducers for~$\G$.
\begin{itemize}
\item[($i$)]
If $\set{A_\agi (\strat^{j\mathsf{K}}_\agi)}_{\agi \in \Agt}$ 
is winning for~$\reach$ in~$\G$,
and $\G^\mathsf{K}$ fulfills the PDK condition, 
then $\set{\alpha^{j\mathsf{K}}_\agi}_{\agi \in \Agt}$ is winning 
for~$\reach^{j\mathsf{K}}$ in~$\G^{j\mathsf{K}}$.
\item[($ii$)]
If $\set{\alpha^{j\mathsf{K}}_\agi}_{\agi \in \Agt}$ is winning 
for~$\reach^{j\mathsf{K}}$ in~$\G^{j\mathsf{K}}$, 
then $\set{A_\agi (\strat^{j\mathsf{K}}_\agi)}_{\agi \in \Agt}$ 
is winning for~$\reach$ in~$\G$.
\end{itemize}
\end{theorem}

\begin{proof}
The proof adapts the one of Theorem~\ref{thm:strategy-correspondence},
but refers now to Lemma~\ref{lem:iMKBSCCharacterisation} to relate
the plays in~$\G$ with those in~$\G^{j\mathsf{K}}$. 

($i$)
Let $\set{A_\agi (\strat^{j\mathsf{K}}_\agi)}_{\agi \in \Agt}$ 
be winning for~$\reach$ in~$\G$,
and let $\G^\mathsf{K}$ fulfill the PDK condition.
Let $\pi^{j\mathsf{K}} = s_0 \sigma_0 s_1 \sigma_1 \ldots$ be an 
arbitrary outcome of 
$\set{\alpha_\agi^{j\mathsf{K}}}_{\agi \in \Agt}$ in~$\G^{j\mathsf{K}}$.
Then, by Lemma~\ref{lem:iMKBSCCharacterisation} and
Lemma~\ref{lem:PDKforGKK}, and since 
$G^\mathsf{K}$ fulfills the PDK condition, the sequence
$\pi = l_0 \sigma_0 l_1 \sigma_1 \ldots$ such that 
$\set{l_k} = \Cap s_k$ for all $k \geq 0$, 
must be a full play  in~$\G$. 
Now, by Definition~\ref{def:generalised-induced-transducer}, 
$\pi$ must be an outcome of
$\set{A_\agi (\strat^{j\mathsf{K}}_\agi)}_{\agi \in \Agt}$ in~$\G$. 
Since $\set{A_\agi (\strat^{j\mathsf{K}}_\agi)}_{\agi \in \Agt}$ 
is winning for~$\reach$ in~$\G$,
$\pi$ must be winning for~$\reach$ in~$\G$,
and hence, by the definition of~$\reach^{j\mathsf{K}}$, 
$\pi^{j\mathsf{K}}$ must be winning for~$\reach^{j\mathsf{K}}$ 
in~$\G^{j\mathsf{K}}$.
But~$\pi^{j\mathsf{K}}$ is arbitrary, and therefore 
$\set{\alpha^{j\mathsf{K}}_\agi}_{\agi \in \Agt}$ must be winning 
for~$\reach^{j\mathsf{K}}$ in~$\G^{j\mathsf{K}}$. 

($ii$)
Let $\set{\alpha^{j\mathsf{K}}_\agi}_{\agi \in \Agt}$ be winning 
for~$\reach^{j\mathsf{K}}$ 
in~$\G^{j\mathsf{K}}$. Let $\pi = l_0 l_1 l_2 \ldots$ be an 
arbitrary outcome 
of $\set{A_\agi (\strat^{j\mathsf{K}}_\agi)}_{\agi \in \Agt}$ 
in~$\G$. This
outcome induces a corresponding sequence of joint observations, 
from which,
using $\set{A_\agi (\strat^{j\mathsf{K}}_\agi)}_{\agi \in \Agt}$, 
one can recover the corresponding sequence of joint actions. By 
Lemma~\ref{lem:iMKBSCCharacterisation}, these two sequences (of 
joint actions and joint observations) give rise to a unique 
play $\pi^{j\mathsf{K}} = s_0 \sigma_0 s_1 \sigma_1 \ldots$ 
in~$\G^{j\mathsf{K}}$ such that $\set{l_k} = \Cap s_k$ for all 
$k \geq 0$. 
Now, by Definition~\ref{def:generalised-induced-transducer}, 
$\pi^{j\mathsf{K}}$ must be an outcome of
$\set{\alpha^{j\mathsf{K}}_\agi}_{\agi \in \Agt}$ in~$\G^{j\mathsf{K}}$. 
Since $\set{\alpha^{j\mathsf{K}}_\agi}_{\agi \in \Agt}$ is winning 
for~$\reach^{j\mathsf{K}}$ in~$\G^{j\mathsf{K}}$, $\pi^{j\mathsf{K}}$ 
must be winning for~$\reach^{j\mathsf{K}}$ in~$\G^{j\mathsf{K}}$,
and hence, by the definition of~$\reach^{j\mathsf{K}}$, $\pi$ must 
be winning for~$\reach$ in~$\G$. But~$\pi$ is arbitrary, and
therefore $\set{A_\agi (\strat^{j\mathsf{K}}_\agi)}_{\agi \in \Agt}$ 
must be winning for~$\reach$ in~$\G$. 
\end{proof}

\subsection{Generalised knowledge representation}
\label{subsec:generalised-knowledge-representation}

Let us denote by~$A^{(j+1)}$ the domain of knowledge states that
can arise in the game structures $(\G^{j\mathsf{K}}|_\agi)^\mathsf{K}$,
and by~$B^{(j)}$ the ones that can arise in~$\G^{j\mathsf{K}}$. 
The elements of~$A^{(j+1)}$ are non-empty sets of elements from~$B^{(j)}$,
and the sets~$A^{(j)}$ can thus be defined formally, for $j>0$, 
by:
$$ A^{(j+1)} \defeq 2^{B^{(j)}} $$
The elements of~$B^{(j)}$ are tuples of elements of~$A^{(j)}$,
one for each agent $\agi \in \Agt$, and thus, the sets~$B^{(j)}$ 
can be defined inductively as follows: 
$$ \begin{array}{rcll}
    B^{(0)}  &  \defeq  &  L &  \\
    B^{(j)}  &  \defeq  &  (A^{(j)})^\Agt & ~~~~~\mbox{for $j>0$}
    \end{array} $$
So, as a structure for representing the \defstyle{$j$-order knowledge} 
of agents we will use elements of~$A^{(j)}$. 

As already mentioned, the elements of~$A^{(j+1)}$ that can actually
arise from the iterated MKBSC have the property that they are 
sets of tuples from $(A^{(j)})^\Agt$ that agree on their $\agi$-th 
component. 
This observation suggests that there are more compact and meaningful 
representations of the knowledge structures, as we will now show. 

\paragraph{Knowledge trees}
For teams of two agents, the knowledge states in the MKBSC-iterated games 
can be represented as pairs of \textbf{knowledge trees}, one for each 
agent, of depth being the iteration index. 
For example, consider the following knowledge state in~$\G^{2\mathsf{K}}$ 
from our running example:
$$ (\set{(\set{\mathsf{good}}, \set{\mathsf{bad}, \mathsf{good}})}, 
    \set{(\set{\mathsf{bad}}, \set{\mathsf{bad}, \mathsf{good}}), \  (\set{\mathsf{good}}, \set{\mathsf{bad}, \mathsf{good}})}) $$
By factoring out the common $\agi$'th component for each agent~$\agi$,
this structure can be equivalently represented as a pair of trees:

\begin{center}
\includegraphics[scale=.3]{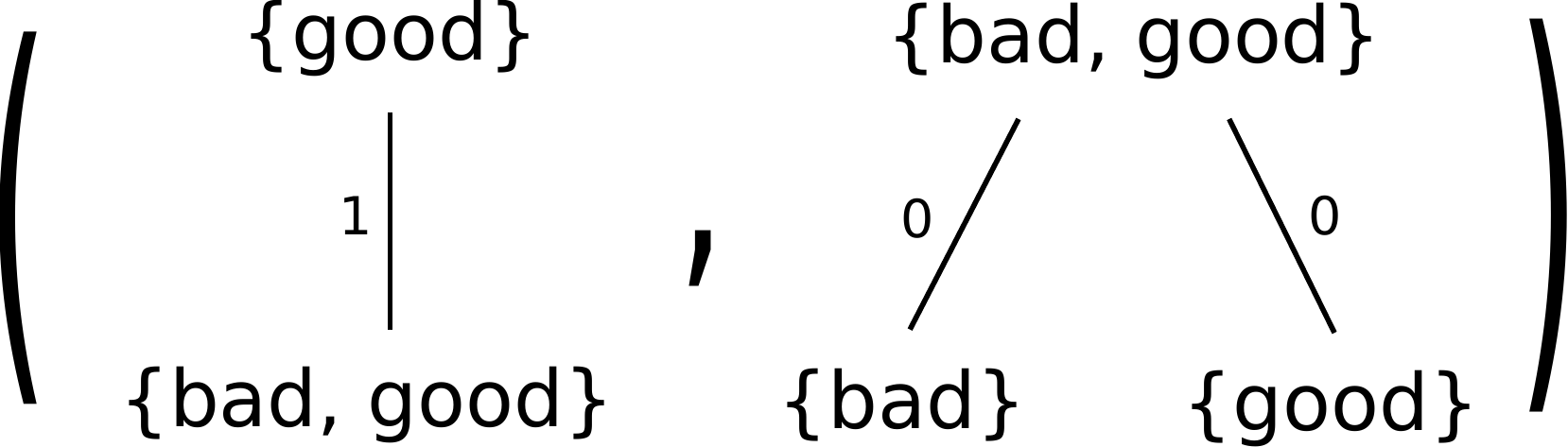}
\end{center}

\noindent
These trees represent the \emph{second-order knowledge} of the two
robots: 
robot~$0$ knows that the grip is good, but that  
robot~$1$ is uncertain about whether the grip is good or bad, while 
robot~$1$ is uncertain about whether the grip is good or bad, but 
knows that robot~$0$ knows which of the two is the case. 

\begin{figure}[ht]
    \centering
    \includegraphics[scale=.12]{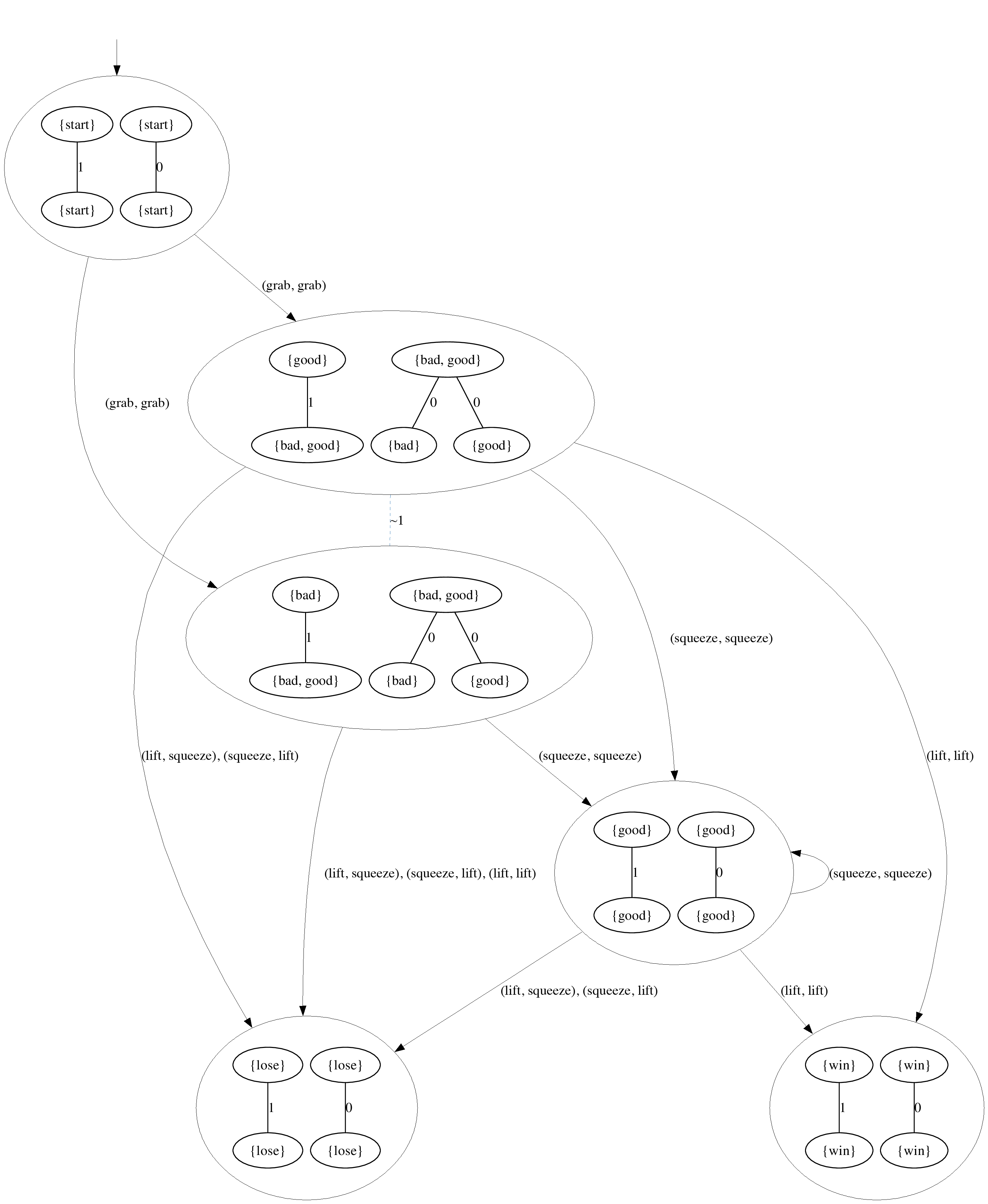}
    \caption{The game $\G^{2\mathsf{K}}$ with explicit knowledge trees.}
    \label{fig:rep-prod-trees}
\end{figure}

A visualisation of~$\G^{2\mathsf{K}}$ for our running example, with 
states represented by the respective pairs of knowledge trees is
shown on Figure~\ref{fig:rep-prod-trees}. It has been obtained 
by a modified version of our tool\footnote{Described in~\cite{han-ros-19-bsc} and available from \texttt{github.com/larasik/mkbsc}.}.
\begin{figure}[ht]
    \centering
    \includegraphics[scale=.15]{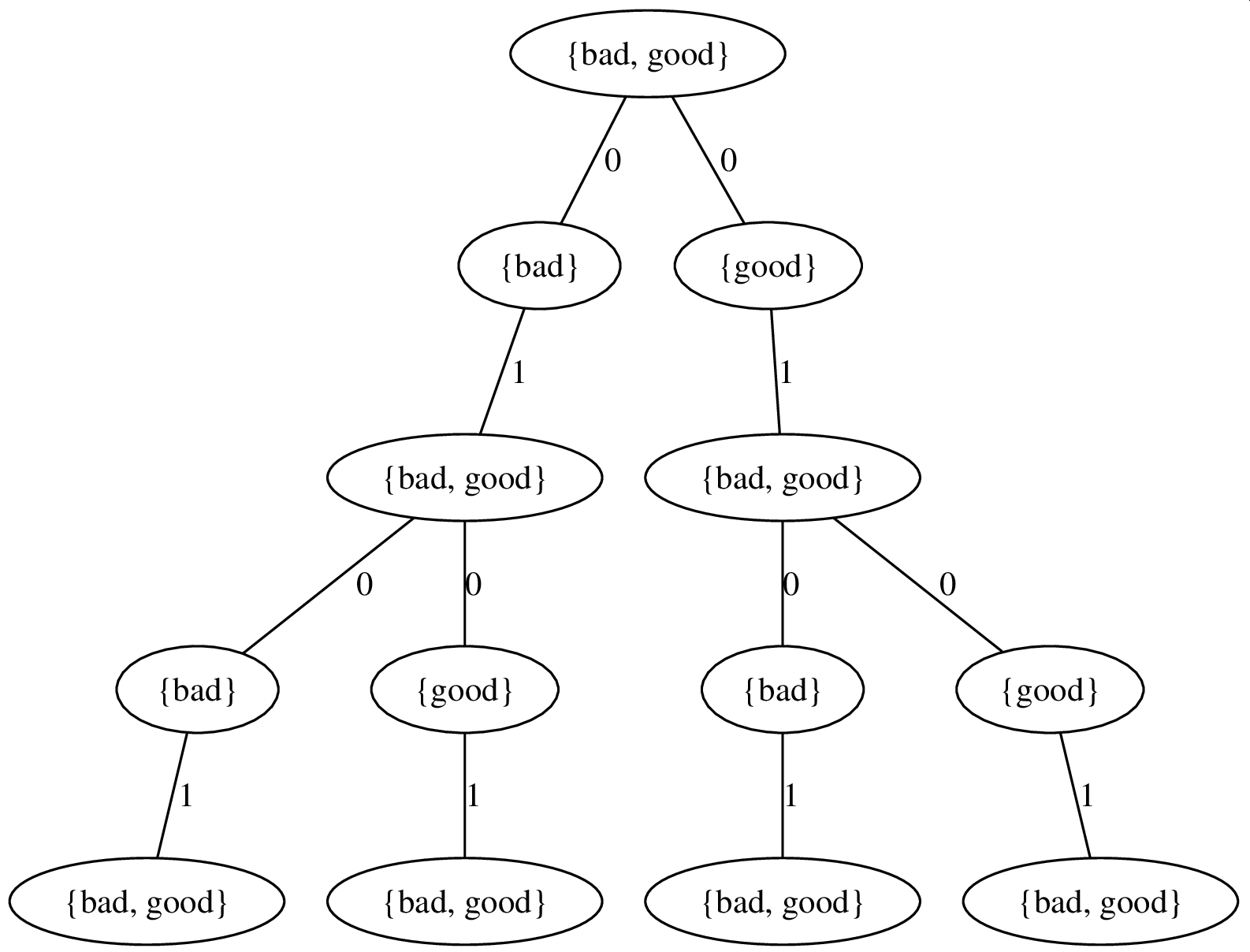}
    \caption{A knowledge tree from $\G^{5\mathsf{K}}$.}
    \label{fig:five-tree}
\end{figure}
Knowledge trees can have any finite depth. For instance, a knowledge 
tree of robot~1 from $\G^{5\mathsf{K}}$ is depicted on 
Figure~\ref{fig:five-tree}.

Representing knowledge in the form of trees, rather than as recursive 
tuples of sets of locations, can play an important role in 
\emph{explaining to humans} knowledge-based strategies that have been 
synthesised algorithmically with our 
method\footnote{%
At the same time it should be noted that such trees are more 
efficiently represented by DAGs.}.

Our notion of knowledge trees is akin to 
the notion of $k$-trees of~\cite{DBLP:journals/iandc/Meyden98}. 
The latter notion is finer than ours in that, in a $k$-tree,
every node (i.e., set) is connected to a particular element
of the parent node (set) rather than with the whole set 
(as it is the case in our representation). 
Some details on $k$-trees can be found in~\ref{app:k-trees},
where we have slightly modified the original presentation,
adapting it to the current set-up. 
Technically and conceptually similar (yet, not equivalent) are 
the $k$-worlds in~\cite{Fagin1991}.  

Lastly, analogous structures have also been used 
in decision and game theory, in the context of multi-player decision making, to represent the knowledge of players who use mental models of the other players when deciding how to act, see e.g.\ the "level-$k$ types" in~\cite{STAHL1995218}.

\subsection{Generalised knowledge update}
\label{subsec:generalised-knowledge-update}

For $j \geq 0$,
taking $A^{(j+1)}$ as the $(j+1)$-order knowledge representation, every 
individual memoryless strategy $\strat^{(j+1)\mathsf{K}}_\agi$
in $(\G^{j\mathsf{K}}|_\agi)^\mathsf{K}$
can be seen as an individual $(j+1)$-order knowledge-based strategy for 
agent~$\agi$ in~$\G$. 

\begin{definition}[Generalised Knowledge Update]
\label{def:gen-knowledge-update}
For $j>0$, $s \in A^{(j+1)}$, $\act_\agi \in \Act_\agi$ and 
$o_\agi \in \Obs_\agi$, the \textbf{generalised knowledge update}
function of agent~$\agi \in \Agt$ is defined inductively as follows: 
$$ \delta^{j+1}_\agi (s, \act_\agi, o_\agi) \defeq 
   \setdef{(\delta^j_{\agi'} (t (\agi'), \act (\agi'), 
                              o (\agi')))_{\agi' \in \Agt}}
       {\begin{array}{l} 
          t \in s, \act \in \Act, o \in \Obs^p,  \\
          \act(\agi) = \act_\agi, o (\agi) = o_\agi
        \end{array}} $$
from which the inconsistent tuples are pruned out 
(as in the Composition step of Definition~\ref{def:mkbsc}), and
where the base case $\delta^1_\agi \defeq \delta_\agi$ is as 
given in Definition~\ref{def:knowledge-update}. 
\end{definition}

As the following result shows, generalised knowledge 
update~$\delta^{j+1}_\agi$ is essentially knowledge 
update~$\delta_\agi$ applied on $\G^{j\mathsf{K}}$
(instead of on~$\G$).
To formalise this, one has to realise that agent 
observations in the original game~$\G$ relate to
observations in the expanded games much in the same
way as the locations relate to knowledge states
(see page~\pageref{def:hat-s}).
For an observation $o_\agi \in \Obs_\agi$, let
$o^j_\agi$ denote the set 
$\setdef{s \in B^{(j)}}{\hat{s} (\agi) \subseteq o_\agi}$.

\begin{lemma}
\label{lem:relating-deltas}
Let $j \geq 1$, $s \in A^{(j+1)}$, $\act_\agi \in \Act_\agi$ and 
$o_\agi \in \Obs_\agi$.
Then:
$$ \delta^{j+1}_\agi (s, \act_\agi, o_\agi) =
   \setdef{l' \in o^j_\agi}
      {\exists \act \in \Act.\ (\act (\agi) = \act_\agi 
       \wedge \exists l \in s.\ (l, \act, l') \in \Delta^{j\mathsf{K}})} $$
\end{lemma}

\begin{proof}
By mathematical induction on~$j$. 
The base case of $j=1$ is established in the proof of
Theorem~\ref{thm:strategy-equivalence}. 
Assume (as the induction hypothesis) that the result holds for~$j$.
We show that the result then follows for~$j+1$.

Let $s \in A^{(j+1)}$, $\act_\agi \in \Act_\agi$, 
$o_\agi \in \Obs_\agi$, and let $o^j_\agi$ 
be as defined above.
Then:
$$ \begin{array}{cll}
   & \delta^{j+1}_\agi (s, \act_\agi, o_\agi) & \\
   = & 
     \setdef{(\delta^j_{\agi'} (t (\agi'), \act (\agi'), 
                              o (\agi')))_{\agi' \in \Agt}}
       {\begin{array}{l} 
          t \in s, \act \in \Act, o \in \Obs^p,  \\
          \act(\agi) = \act_\agi, o (\agi) = o_\agi
        \end{array}} & 
     \{\mbox{Def.~\ref{def:gen-knowledge-update}}\} \\
   = & 
     \begin{array}{l}
     \left\{ (\setdefn{l' \in o^{j-1}_{\agi'}}
        {\exists \act \in \Act.\ (\act (\agi') = \act_{\agi'} 
            \wedge \exists l \in t (\agi').\ (l, \act, l') \in 
                   \Delta^{(j-1)\mathsf{K}})})_{\agi' \in \Agt} \right. \\
     \>\>\> \left| \>\> t \in s, \act \in \Act, o \in \Obs^p, 
               \act(\agi) = \act_\agi, o (\agi) = o_\agi \right\}
     \end{array} & 
     \{\mbox{Ind.hyp.}\} \\
   = & 
     \setdef{l' \in o^j_\agi}
      {\exists \act \in \Act.\ (\act (\agi) = \act_\agi 
       \wedge \exists l \in s.\ (l, \act, l') \in \Delta^{j\mathsf{K}})} & 
     \{\mbox{Def.~\ref{def:mkbsc}}\} \\
   \end{array} $$
\end{proof}

The following result generalises Theorem~\ref{thm:strategy-equivalence} 
to the iterated MKBSC: the strategy profiles resulting from the translation 
to transducers and to knowledge-based strategies are equivalent, i.e., 
give rise to the same sets of outcomes.

\begin{theorem}[Generalised Strategy Equivalence]
\label{thm:gen-strategy-equivalence}
Let~$\G$ be a \magii for a set of agents~$\Agt$, let $j \geq 1$, 
and let $\G^{j\mathsf{K}}$ be the $j$-iterated MKBSC expansion of~$\G$. 
Let $\set{\alpha^{j\mathsf{K}}_\agi}_{\agi \in \Agt}$ be a profile 
of observation-based memoryless strategies in~$\G^{j\mathsf{K}}$, and
$\set{A_\agi (\strat^{j\mathsf{K}}_\agi)}_{\agi \in \Agt}$ be the 
corresponding profile of induced transducers for~$\G$. 
Then, the strategy profile 
$\set{A_\agi (\strat^{j\mathsf{K}}_\agi)}_{\agi \in \Agt}$,
and the profile of $j$-order knowledge-based strategies based on 
$\set{\alpha^{j\mathsf{K}}_\agi}_{\agi \in \Agt}$ and 
$\set{\delta^j_\agi}_{\agi \in \Agt}$, give rise to the same
set of outcomes in~$\G$.
\end{theorem}

\begin{proof}
We show that
$\tau^j_\agi (s, o_\agi) = 
 \delta^j_\agi (s, \alpha^{j\mathsf{K}}_\agi (s), o_\agi)$
for all $j \geq 1$, $s \in A^{(j+1)}$ and $o_\agi \in \Obs_\agi$.
The result then follows from 
Definition~\ref{def:generalised-induced-transducer},
Definition~\ref{def:finite-memory-strategy}, and the
definition of $j$-order knowledge-based strategies.

The case when $j=1$ is established by 
Theorem~\ref{thm:strategy-equivalence}.
Let $j \geq 1$, $s \in A^{(j+1)}$, $\act_\agi \in \Act_\agi$, 
$o_\agi \in \Obs_\agi$, and let $o^j_\agi$ 
be as defined above.
Then:
$$ \begin{array}{cll}
   & \tau^{j+1}_\agi (s, o_\agi) & \\
   = & 
     \mbox{the unique $s' \in S_\agi$ such that $\hat{s}' \subseteq o_\agi$ 
           and $(s, \strat^{(j+1)\mathsf{K}}_\agi (s), s') \in 
           \Delta^{(j+1)\mathsf{K}}_\agi$} &
     \{\mbox{Def.~\ref{def:generalised-induced-transducer}}\} \\
   = &
     \setdef{l' \in o^j_\agi}
        {\exists l \in s.\ (l, \alpha^{(j+1)\mathsf{K}}_\agi (s), l') 
           \in \Delta^{j\mathsf{K}}_\agi} &
     \{\mbox{Def.~\ref{def:mkbsc}.2}\} \\
   = &
     \setdef{l' \in o^j_\agi}
        {\exists \act \in \Act.\ (\act (\agi) = \alpha^{(j+1)\mathsf{K}}_\agi (s) 
         \wedge \exists l \in s.\ (l, \act, l') \in \Delta^{j\mathsf{K}})} &
     \{\mbox{Def.~\ref{def:mkbsc}.1}\} \\
   = &
     \delta^{j+1}_\agi (s, \alpha^{(j+1)\mathsf{K}}_\agi (s), o_\agi) &
     \{\mbox{Lem.~\ref{lem:relating-deltas}}\} \\
   \end{array} $$
if such an~$s'$ exists; otherwise, by 
Lemma~\ref{lem:iMKBSCCharacterisation}, 
$\delta^{j+1}_\agi (s, \alpha^{(j+1)\mathsf{K}}_\agi (s), o_\agi)$
is undefined.
\end{proof}

As we showed in Lemma~\ref{lem:PDKforGKK}, for all $j>1$, the 
expansions~$\G^{j\mathsf{K}}$ satisfy the PDK condition.
By composing our results we obtain that, for all $j>1$, our strategy 
synthesis method is \emph{complete} for the class of $j$-order knowledge-based  strategies with respect to observable reachability objectives whenever~$\G^\mathsf{K}$ satisfies the PDK condition, in the sense that if a winning profile of $j$-order knowledge-based strategies exists, it will be found with our method. 

Further, since the observation-based memoryless 
strategies are preserved by the MKBSC, the sets of strategies produced 
by the iterative approaches above
grow monotonically: the class of $j$-order knowledge-based strategies 
subsumes any lower-order class. 
In other words, \emph{increasing the order of knowledge increases the 
strategic ability of the team}.

\section{Follow-up discussion} 
\label{sec:IMKBSC-Follow-up}

In this section we discuss some aspects of the MKBSC that we 
consider of importance, and which will be explored in follow-up work. 

\subsection{Utilising the iterated MKBSC}
\label{subsec:iter-mkbsc-utility}

There are two dual approaches to using the constructions presented here 
for synthesising knowledge-based strategies: 

\begin{itemize}
\item
\textbf{Global: }
Keep applying iteratively the MKBSC on the game graph and then search for an observation-based memoryless strategy profile in the resulting expanded game until -- if ever -- such strategy profile is found. Then convert it to a knowledge-based strategy profile. 

\item
\textbf{Local: }
Keep incrementing the epistemic depth and exploring the reachable part of the game graph produced (at the current depth) by means of the knowledge update functions. Then, search for a memoryless strategy in that graph which, if found, will by construction be a knowledge-based one.
\end{itemize}

\noindent As indicated earlier, these two approaches are equivalent.

\subsection{Stabilisation of the Iterated MKBSC}  
\label{subsec:IMKBSC-Stabilisation}

For some \magii games the iterated construction eventually stabilises, in the sense that it starts producing isomorphic games, even though the internal structure of the locations (the tuples of knowledge trees) grows unboundedly. For instance, for the game graph of our running example, $\G^{3\mathsf{K}}$ is isomorphic to~$\G^{2\mathsf{K}}$. 
We then say that game graph~$\G^{2\mathsf{K}}$ is \emph{stable}.

Thus, the global approach outlined above can be augmented with a check for
stabilisation.
Furthermore, as explained in the end of Section~\ref{sec:IteratedMKBSC}, if we know that the construction will stabilise, we may directly iterate until that point and only search for memoryless strategies in the stable game (since we are guaranteed to not miss any such strategy that would have been found in some of the intermediate games).

This leads to a number of interesting questions about stabilisation and the properties of stable game graphs: 

\begin{itemize}
\item 
To begin with, what does stabilisation mean in terms of the knowledge encoded in the states of stabilised games, and in terms of existence of knowledge-based strategies in them? 

We only note here that stabilisation corresponds to the existence of a
finite knowledge representation that contains the higher-order 
knowledge of the agents of \emph{any} order.
The knowledge encoded in the locations of stable game graphs could thus be ``folded'' into recursive representations, and these eventually allow to capture \textit{common knowledge}. 
It is well-known that coordinated action in multi-agent games 
requires common knowledge
(see, e.g.,~\cite{DBLP:journals/jacm/HalpernM90}).
In the context of our running example, it is worth noting that 
the knowledge state just above the two knowledge states at the 
bottom of Figure~\ref{fig:rep-prod-trees}, while representing
second-order knowledge of the two robots, can also 
be seen, since $\G^{2\mathsf{K}}$ is stable, as representing 
the common knowledge of the robots that they 
both have a good grip. This, in fact, is what justifies that they 
can simultaneously lift at this point.

\item 
What structural conditions on a game are necessary, or sufficient, 
for its iterated MKBSC to eventually stabilise? 

\item 
For which classes of objectives does it suffice to search for memoryless strategies in their stable expansions?  

\item What are the implications of non-stabilisation? 
\end{itemize}

\noindent
These questions will be explored in follow-up work.

\subsection{Limitations of the MKBSC} 
\label{subsec:mkbsc-limitations}

As stated in the Introduction, the chosen knowledge representation is just one possible choice, albeit one with a good intuitive justification. As the following example illustrates, however, it is not the case that whenever there is a profile of observation-based finite-memory strategies that is winning for a given reachability objective, then there is also a winning profile of $j$-order knowledge-based strategies (of the type studied here) for some~$j$.

\begin{figure}[ht]
    \centering
    \includegraphics[scale=.35]{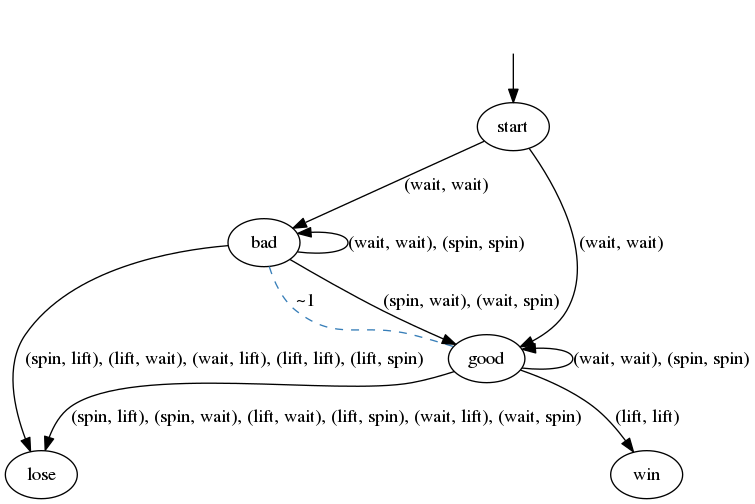}
    \caption{A stable game.}
    \label{fig:cup-turn}
\end{figure}

\begin{example}
\label{ex:limitations}
Consider the game shown in Figure~\ref{fig:cup-turn}.
It models a similar scenario as our cup-lifting game, but now the cup
needs to be oriented suitably before it can be lifted.
The game is stable, meaning that whatever can be achieved by a profile of $j$-order knowledge-based strategies for any $j \geq 1$, it can also be achieved by a profile of observation-based memoryless strategies.
However, for this game, there is no winning profile of observation-based memoryless strategies, while there clearly is a winning profile of observation-based finite-memory ones: after the initial wait, robot~1 waits once more, while robot~0 spins the cup if needed; then the two robots lift the cup.
\end{example}

\section{Related work} 
\label{sec:related-work}

As pointed out in the introduction,  the present work relates in more, or less, essential ways with several major research areas, incl., decentralised cooperative decision making, multi-agent planning, knowledge-based programs, games with imperfect information and strategy synthesis in them, etc., where variations of the general problem in focus of this paper have been explored, some of them quite extensively.
Here we provide a reasonably detailed, yet inevitably incomplete list of the conceptually and technically closest  to our work research areas and topics, 
with some relevant key references on them.

\subsection{Games with imperfect information and knowledge-based strategies} 
\label{subsec:related-work-Games with imperfect information}

Most of the basic notions and components (incl. terminology and notation) of our framework 
 are adopted from general studies of games with imperfect information, such 
as~\cite{DBLP:conf/focs/PetersonR79,DBLP:journals/jcss/Reif84,DBLP:conf/focs/PnueliR90,DBLP:books/cu/11/0001R11}. 

In particular, the \emph{knowledge-based subset construction} (KBSC) for single-agent 
games against Nature has been introduced and studied 
in~\cite{DBLP:journals/jcss/Reif84,DBLP:journals/lmcs/RaskinCDH07}.
The generalisation MKBSC for \magii explored here was first proposed 
in~\cite{lun-17-msc} by the third co-author. Our presentation of the MKBSC is
equivalent to that one, but makes explicit the different stages of the
construction (the original proposal defines directly the expanded game).
This ``deconstruction'' of the original definition has allowed us to 
define a translation of strategies to transducers 
(Definition~\ref{def:induced-transducer}),
and to propose a heuristic for strategy synthesis based on single-agent
games with perfect information (Section~\ref{subsec:strategy-synthesis}).
Furthermore, 
\cite{lun-17-msc} did not provide any formal
characterisation of the construction, but discovered the phenomenon of
stabilisation and made the observation that iterating the MKBSC in effect computes higher-order
knowledge, illustrating this on the cup-lifting game. Finally,
\cite{lun-17-msc} 
made the observation of the limitation of the construction 
discussed here in Section~\ref{subsec:mkbsc-limitations}.

The notion of knowledge-based strategies proposed and explored here
is closely related to the notion of \emph{knowledge-based programs}, 
introduced and studied, e.g., in   
\cite{Fagin1991,DBLP:journals/dc/FaginHMV97,FHMV}.
Our presentation is more abstract and less algorithmic, but our results
can easily be adapted to the latter notion. These may be useful 
for the synthesis of protocols and for programming intelligent agents. 

Constructions representing and using agents' higher-order knowledge have 
been proposed in~\cite{Fagin1991,DBLP:journals/iandc/Meyden98,Berwanger2010,DBLP:conf/fsttcs/BerwangerKP11}. 
All these are essentially related to our MKBSC construction. 
The notion of $k$-trees proposed in~\cite{DBLP:journals/iandc/Meyden98}
was already discussed in 
Section~\ref{subsec:generalised-knowledge-representation}. 
Among the constructions, special mention deserves the \emph{epistemic 
unfolding} introduced and studied in~\cite{DBLP:conf/fsttcs/BerwangerKP11}.
The construction essentially translates a \magii, in a strategy-preserving
fashion, to a single-agent game with perfect information. The resulting 
(expanded) game, however, is  generally infinite (even when collapsed
to so-called ``homomorphic cores''), and the construction is thus not 
guaranteed to terminate. 
In contrast, our MKBSC construction always results in a finite game, 
but does not necessarily remove the imperfect information. Another difference 
is that the unfolding is based on an ``epistemic model'' common to all agents, 
and thus addresses the (YN) case discussed in 
Section~\ref{subsec:KnowledgeBasedStrategies}.

\subsection{Logics for multi-agent knowledge and strategic reasoning} 
\label{subsec:related-work-Logics for MAS}

While we have not involved formal logical languages and systems in this study, we note that it is essentially related to various logics  and models of multi-agent knowledge and strategic reasoning. 

First, there is a clear link of our work with \emph{multi-agent epistemic logic}. Every agent in our framework is implicitly assumed to be a perfect (ideal) reasoner and its knowledge satisfies the standard S5 principles. Thus, the knowledge, both factual and higher-order, of all agents is modelled in multi-agent epistemic frames by epistemic indistinguishability relations \cite{DitmarschvdHoekKooi}, which are equivalence relations that partition the state space into families of information/knowledge states (see e.g.\  \cite{FHMV}).  
These are mutually inter-definable with \emph{Aumann structures}, and higher-order knowledge of the agents can be computed in either of them in a standard way (see again~\cite{FHMV}). These epistemic structures naturally arise in the \magii models, but we do not define and deal with them explicitly, since we do not need to involve  explicitly epistemic logic, or any other formal logical system in the present work. 

Furthermore, there are 2 hierarchies of epistemic structures naturally arising in  \magii models and the problems we study. The first one is the hierarchy of \emph{static higher-order knowledge}, associated with the hierarchy of iterations of the MKBSC construction applied to the original game. Every iteration increases the order of explicitly represented agents' knowledge, starting with 0-order (the factual knowledge about the current state of the game), then 1-order (the knowledge about the other agents' 0-order knowledge), 2nd order, etc. This is static knowledge because it is not associated with any particular play. The second one is the hierarchy of \emph{(0-order) dynamic knowledge}, associated with each particular play in the given model, where the agents re-compute it after every transition and new observation they make, in the way described in this paper. Here we also mention the close conceptual links with  \emph{dynamic epistemic logic (DEL)} \cite{DitmarschvdHoekKooi}, esp. the mechanism of epistemic updates of the knowledge representing models. Some recent studies, relating DEL and its use in solving concurrent games, include \cite{DBLP:conf/ijcai/MaubertPSS20,DBLP:conf/ijcai/MaubertPS19}, technical ideas in which can be used for further extension of the present work to a non-cooperative setting. 

These two hierarchies of knowledge can be interleaved in a natural way, by producing a combined hierarchy of \emph{higher-order dynamic knowledge}, the capturing and utilisation of which for synthesising winning strategy profiles for the team of agents we have explored here. We leave the study of that combined hierarchy from the perspective of multi-agent epistemic logic to future work, but we note its relationship with the hierarchy of $k$-trees, defined and explored originally in~\cite{DBLP:journals/iandc/Meyden98}, and further works, 
including~\cite{DBLP:conf/aaai/HuangM12,DBLP:journals/corr/MeydenV13}. 

Another natural link can also be made with general game description languages (GDL) as a different framework for knowledge representation in games, and \cite{DBLP:journals/ai/EngesserMNT21} provides a  comparative study of the GDL approach with the DEL-based framework that can also be useful for further work on the topic. 

An important related line of research on \emph{models and logics for games with (perfect and) imperfect information} comprises a large number of papers, going back to \cite{vdHW04,Jamroga04ATEL,Jamroga07constructive-jancl}, focusing mainly on the logical properties of semantics, expressiveness, deciding satisfiability, etc. 
More recent and also more computationally focused works in that line of research include: 
\begin{itemize}
\item  \cite{DBLP:journals/jancl/GuelevDE11}, developing a model checking algorithm for a variant of the alternating time temporal logic ATL with knowledge operators, assuming incomplete information and perfect recall, but also communication between the agents, so the strategies considered there employ the distributed knowledge of the agents. 

\item \cite{DBLP:journals/ai/Huang15}, which develops (bounded) model checking methods for reachability and other problems, expressible in the logic ATL, under the assumption of imperfect information and perfect recall, but with bounded horizon.  

\item \cite{DBLP:conf/atal/JamrogaMM19} which considers a variant of the logic NatATL (``ATL with Natural Strategies") introduced in  \cite{DBLP:journals/ai/JamrogaMM19} with imperfect information and studies the model checking problem for it. In this logic, bounds are imposed on the complexity of the admissible strategies, assuming that they are represented by lists of guarded actions.
\end{itemize}
More related references can be found 
in~\cite{AgotnesGorankoJamrogaWooldridge-HEL15}. 

The closest link of our study with that research line are the concurrent game models with incomplete and imperfect information that are essentially used as models in our work, too. The major distinctions in our study as compared to these are, first, that all agents work as a team; second, that there is a non-deterministic environment deciding the outcome of the action profiles of the team; third, that we do not employ (yet) a formal logical language here, and fourth, that the strategy profile is designed externally and then communicated, either fully, or only locally, to the individual agents. These make the methods and results of our work substantially different from those presented in the literature mentioned above.

\subsection{Decentralized partially observable Markov decision processes} 
\label{subsec:related-work-Dec-POMDP}

Technically, our framework of \magii models is a special case of \emph{decentralized partially observable Markov decision processes (Dec-POMDP)}  
\cite{DBLP:journals/mor/BernsteinGIZ02,DBLP:journals/aamas/SeukenZ08,DBLP:books/sp/12/Oliehoek12,DBLP:series/sbis/OliehoekA16}, modelling multi-agent planning and decision-making under uncertainty, where the policy planning is centralized whereas the execution is assumed decentralized because of the lack of (adequate)  communication between the agents in the execution phase. 
The essential differences of the general framework of Dec-POMDPs from our 
\magii models are as follows: 

\begin{itemize}
\item The transitions in Dec-POMDPs are determined by transition probability functions with explicitly specified distributions associated with each action profile, whereas in \magii models they are randomly settled for each action profile by Nature. (The possible outcomes can be assumed uniformly distributed, but without using that assumption for optimising the team's joint policy.)     

\item The \emph{reward function} in Dec-POMDPs is typically quantitative and the aim is to maximize it, 
whereas in \magii models it is a qualitative objective, typically reachability or safety.  

\item The agents' observations  in Dec-POMDPs are stochastic, subject to given probability distributions, 
whereas in \magii models they are deterministically determined by the states. 

\item  A finite  \emph{horizon} (number of transition steps for optimising the reward) is usually  explicitly assumed. On the other hand, the horizon on the problem that we study here is implicitly assumed unbounded.  
\end{itemize}

The typical decision problem studied for Dec-POMDPs is as follows: given a Dec-POMDP $M$, a positive integer horizon $T$ and a reward threshold $K$, the question is whether there is a joint policy for all agents that yields a total reward in $T$ steps which is at least $K$. This problem is clearly decidable and the main research problem in the studies cited above is to analyse and determine its complexity under various additional assumptions. Typically, it is NEXPTIME-complete, even in the 2-agent case. 
This is where the main difference with our study occurs. The reward function in our framework is very simple: it assigns a reward $1$ if the objective is reached, otherwise assigns 0, and we do not assume a pre-defined finite horizon, which makes the respective reachability problem in the focus of our study generally undecidable, and only semi-decidable -- just like the infinite-horizon Dec-POMDP problems under various optimality criteria, cf. \cite{DBLP:journals/mor/BernsteinGIZ02}.  
Respectively, our main aim is to develop semi-decision procedures for constructing successful strategy (policy) profiles, and a major problem of our study is to ensure termination of these procedures. And, very importantly, we essentially  employ the higher-order knowledge of the agents in the design of these strategy profiles, which is (at least explicitly) not taken into account in the alternative approaches and methods mentioned further. 

From the large body of literature on Dec-POMDPs, we have identified the following directions and works as the closest to our study: 

\begin{itemize}

\item 
Approximation algorithms for solving the infinite-horizon problems in Dec-POMDPs with quantitative reward functions have been proposed, e.g.\  in~\cite{DBLP:conf/ijcai/BernsteinHZ05}, using a joint controller with a correlation device that sends signals to all agents, plus a bounded policy iteration algorithm for improving the agents' finite-state controllers; followed by \cite{DBLP:journals/jair/BernsteinAHZ09} where an optimal policy iteration algorithm
for solving DEC-POMDPs is developed, using stochastic finite-state controllers to represent
policies. Other works in this direction include~\cite{DBLP:conf/ecml/SzerC05}, presenting a best-first search algorithm for computing an optimal policy vector, and~\cite{DBLP:conf/uai/AmatoBZ07}, based on a memory-bounded optimization approach using nonlinear optimization techniques.
In~\cite{DBLP:conf/aips/AmatoDZ09}, an incremental policy generation method is applied to Dec-POMDPs with finite horizon using one-step reachability analysis, but the same approach can also be used in infinite-horizon policy iteration. In \cite{DBLP:journals/ijrr/AmatoKACHK16} 
a new MacDec-POMDP planning algorithm is presented that searches over policies represented as finite-state controllers, which can be much more concise and easier to interpret than  representations based on policy trees, and can operate over an infinite horizon. 
In~\cite{DBLP:conf/atal/AmatoZ09},   the infinite-horizon assumption is replaced by indefinite-horizon. 

The methods and results mentioned above, as well as other related works in the area, are well surveyed in \cite{DBLP:journals/aamas/SeukenZ08,DBLP:books/sp/12/Oliehoek12,DBLP:conf/cdc/AmatoCGUK13}.  It is important to emphasize that all these methods are specifically applicable to optimising \emph{quantitative} reward functions, but essentially not -- at least, not naturally and efficiently -- in the case of quantitative reachability objectives studied here.  

\item \cite{DBLP:journals/jair/DibangoyeABC16} proposes transforming a Dec-POMDP into a  
deterministic MDP,  based on ``complete information-states" which represent the joint history of the individual ``decision rules" applied by the agents, starting with a given initial belief state. 
That enables reduction of finding an optimal separable joint policy in the original Dec-POMDP to the same problem in the resulting complete information-state MDP, by applying various methods developed for the latter problem. Again, the methods explored here apply to optimising \emph{quantitative} reward functions.

\item  \cite{DBLP:conf/aaai/BrafmanSZ13} studies the decision problem with finite horizon, but where the objectives are \emph{qualitative}, reachability goals, rather than maximizing quantitative long-term reward functions, as in general Dec-POMDPs. It is shown there that, under certain assumptions (incl. an explicit finite horizon and a shared initial belief state) the complexity of the problem is as hard as the one for standard Dec-POMDP, and a method for computing a solution for the deterministic case based on compilation to classical planning is presented.   

\item 
We also note the recent works \cite{DBLP:conf/aaai/SaffidineSZ18,DBLP:journals/ai/ZanuttiniLSS20} which show, inter alia, that all (deterministic) joint policies for QDec-POMDPs can be (succinctly) represented as multi-agent  knowledge-based programs, but without discussing the question of how such policies (and their representing programs) for a given objective can be constructed. This result is quite close in spirit to our observation that the intensional and extensional views on knowledge-based strategies are equivalent.  

\end{itemize}

\medskip
In \cite{DBLP:journals/jair/PynadathT02} the different, yet quite similar  to Dec-POMDPs   framework of \emph{Multiagent Team Decision Problem (MTDP)} is presented, where possible explicit communication between the agents is also considered, and the assumption of agents' perfect recall for the agents is made. In particular, "state-estimator functions" are added and used to model and update  the current belief states based on the agents' communication and recall. These are conceptually similar to our abstract mechanism for knowledge updates.  Again, only quantitative reward functions are considered in that work, and the complexities of solving the respective decision problems are studied and shown in \cite{DBLP:journals/aamas/SeukenZ08} to be of the same complexity as for Dec-POMDPs. 

\medskip
In summary, none of the works on solving Dec-POMDPs or related problems surveyed above addresses the case of reachability objectives with unbounded horizon studied here, nor do they propose a solution that is adequate and efficient for solving that problem. Thus, whereas technically our work falls in the broader framework of Dec-POMDPs, both the decision problem studied here and our approach to its solution are substantially different from those mentioned above.

\subsection{Planning under uncertainty} 
\label{subsec:related-work-Planning under uncertainty.}

From a more general perspective, our work also relates, albeit in less essential ways, to other formal frameworks and studies of cooperative multi-agent planning under uncertainty 
\cite{DBLP:conf/kr/SardinaGLL06,DBLP:journals/apin/TorrenoOS14,DBLP:conf/aaai/MuiseBFMMPS15,DBLP:journals/csur/TorrenoOKS17} 
and, in particular, multi-agent epistemic planning
\cite{DBLP:journals/jancl/BolanderA11,DBLP:conf/ecai/CooperHMMR16,DBLP:journals/corr/EngesserBMN17,DBLP:conf/lori/LiW19}.
Some of these works, e.g.\ \cite{DBLP:conf/kr/SardinaGLL06}, also assume incompleteness of the information about the domain, viz. that some of that information is unavailable at plan time, but can be acquired at runtime by the agents executing the plan, by also taking into account the higher-order knowledge or beliefs of the agents.

Major differences from most of these works are that in our framework agents are assumed to cooperate but not communicate with each other, their collective goal is not epistemic but ontic, and they act collectively against Nature. Still, some ideas and techniques from these works are quite relevant to our study and worth exploring further.  

We also note the general link with (multi-agent) temporal planning 
\cite{DBLP:conf/ijcai/CushingKMW07,DBLP:conf/aips/CooperMR12,DBLP:journals/jair/CooperMR14}. Although the models and problems studied here, and the method we developed for their solutions, are different from those explored in that area, some ideas and approaches from the latter, such as easy-to-use modelling languages, can be beneficial for further progress on the topic explored here. 

\subsection{Other related topics and works} 
\label{subsec:related-work-Others}

Of the many other related topics and works, we only mention the implicit link of our study to \emph{theories of mind} (see, e.g., \cite{van_de_Pol_2016}), though in our case the ``minds'' of the agents are only represented by their knowledge known to the other agents, and 
possibly by the strategies which they follow, but not by beliefs, 
intentions, and other attitudes. 
Still, we believe that many interesting phenomena of ``mind'', especially
in the context of Artificial Intelligence, can be studied and explored 
in our simplified framework of knowledge.

\section{Conclusion} 
\label{sec:Conclusion}

We have studied 
agents' (first- and higher-order) knowledge representation in multi-agent games with imperfect information against Nature and its use for synthesising knowledge-based strategy profiles for a team of agents by a supervisor, which then provides each agent with their own strategy and lets them play without them being able to communicate with each other.
In particular, we have introduced and studied the generalised knowledge-based construction MKBSC and have established connections between (transducer-based)
finite-memory strategies and knowledge-based strategies in the original game,
and observation-based memoryless strategies in its iterated MKBSC expansions. 

\paragraph{Conclusions}

From our results one can draw the following conclusions.
First, 
higher-order (nested) knowledge can be based on the notion of 
``most precise estimate of the current state-of-affairs''.
The higher the order  (i.e., the nesting depth) of knowledge, 
the higher the strategic abilities of the team. 
Also,
the higher the uncertainty of the agents, i.e., the less they
observe and know, the higher the benefit from nested knowledge. 
Next, 
for the class of knowledge-based strategies considered here 
(i.e., for the proposed notion of knowledge) and the classes
of reachability and safety objectives, for a given 
bound on the nesting depth of knowledge and under the PDK condition, 
we have an \emph{algorithm} for strategy synthesis; without such a 
bound it is only a semi-algorithm.
Then,
there is a \emph{duality} between the extensional and the intensional
views. The former is more suitable for strategy synthesis,
while the latter can be more convenient in the play, and can
also be used to explain the synthesised strategies. 
And finally,
on some games the iterated MKBSC \emph{stabilises}. If there is no
winning memoryless observation-based strategy in the stable
expansion, then, under the PDK condition, 
there is no winning knowledge-based strategy of any
order in the original game. However, there might still
be a winning finite-memory observation-based strategy in the original
game.

\paragraph{Future work}

In future work we plan to
characterise the class of objectives that can be achieved with 
knowledge-based strategies of the type defined here. 
Next, we plan to capture
formally the notion of ``degree of imperfectness'' of information 
and the intuition (expressed in Section~\ref{subsec:strat-pres}) 
that the MKBSC decreases this degree.
We also plan to
study the strategy synthesis problem after relaxing some of the 
assumptions made here, such as the case (YN) when agents are
permitted to know each others strategies, or when agents do have 
some (limited) communication.
Further, we plan to
study in depth the stabilisation phenomenon of the MKBSC and, 
in particular, characterise the structural conditions for 
stabilisation, and investigate the relationship of stable games 
to common knowledge.
Furthermore, we will
explore other knowledge representations, comparing the respective 
classes of objectives that they are sufficient for, and define
the corresponding expansions following the general scheme. 
We will also
design strategy synthesis algorithms and heuristics, and 
investigate their complexity.
Further, we will explore temporal (epistemic) logic as a means
for defining objectives, and epistemic logic as a means for
representing the individual knowledge-based strategies. 
We will also explore the connection of our work to multi-agent 
epistemic planing.
Finally, we plan to
evaluate the practical utility of the strategy synthesis method
proposed here, and investigate in depth potential application areas.




\newpage
\appendix

\section{$k$-trees}
\label{app:k-trees}

Following~\cite{DBLP:journals/iandc/Meyden98},
we define the set of \defstyle{$k$-trees~$\mathcal{T}_k$} and the set of \defstyle{$\agi$-objective $k$-trees~$\mathcal{T}_{k, \agi}$}, for all agents $\agi \in \Agt$. The set of $\agi$-objective $k$-forests is defined as $\mathcal{F}_{k, \agi} \defeq 2^{\mathcal{T}_{k, \agi}}$.

\begin{definition}[$k$-tree]
Let $\G$ be a \magii over the set of locations~$\Loc$, with agents 
$\Agt = \set{1, 2, \ldots, n}$. 
We recursively define the sets:
$$ \begin{array}{rcl}
    \mathcal{T}_0 & \defeq & \Loc~~~~\mbox{(or rather, $\setdef{\tuple{l, \varnothing, \ldots, \varnothing}}{l \in \Loc}$)}  \\
    \mathcal{T}_{0, \agi} & \defeq & \Loc  \\ \\
    \mathcal{T}_{k+1} & \defeq & 
    \setdef{\tuple{l, F_1, \ldots, F_n}}
    {l \in \Loc \mbox{ and }  F_\agi \in \mathcal{F}_{k, \agi}
 \mbox{ for all } \agi \in \Agt.}  \\    
    \mathcal{T}_{k+1, \agi} & \defeq & \setdef{\tuple{l, F_1, \ldots, F_n} \in  \mathcal{T}_{k+1}}{F_\agi = \varnothing}  \\
    \end{array} $$
\end{definition}

\begin{example}
\label{ex:k-tree}
In a 2-agent game against Nature, consider the 2-tree:
$$ t^{(2)}_1 = \tuple{\loc_2, F^{(1)}_1, F^{(1)}_2}$$
where: 
$$ \begin{array}{rcl}
    F^{(1)}_1 & = & \set{\tuple{\loc_1, \varnothing, \set{\loc_1}}, \tuple{\loc_2, \varnothing, \set{\loc_2, \loc_3}}}  \\
    F^{(1)}_2 & = & \set{\tuple{\loc_2, \set{\loc_1, \loc_2}, \varnothing}, \tuple{\loc_3, \set{\loc_3}, \varnothing}}  \\
    \end{array} $$
This 2-tree models a state of affairs in which the game is in location~$\loc_2$, and agent~1 knows (i.e., considers possible, as modelled by~$ F^{(1)}_1$) that the game is either in location~$\loc_1$ or in~$\loc_2$, and that in the former case agent~2 knows that the game is in~$\loc_1$, while in the latter case, in~$\loc_2$ or~$\loc_3$. The knowledge of agent~2 is analogous, as modelled by~$ F^{(1)}_2$.
\end{example}

Next we define, by mutual recursion, a \defstyle{global update} function~$G_k$,
and a family of local update functions~$H_{k, \agi}$:
$$ \begin{array}{rcl}
    G_k & : & \mathcal{T}_k \times \Act \times \Loc \rightarrow \mathcal{T}_k \\
    H_{k, \agi} & : & \mathcal{F}_{k, \agi} \times \Act_\agi \times \Obs_\agi \rightarrow \mathcal{F}_{k, \agi}
    \end{array} $$

\begin{definition}[Knowledge Update]
Let $\G = (\Loc, \init, \Act, \Delta, \Obs)$ be a \magii. 
We recursively define the functions: \\

$ G_0 (l, \act, l')  \defeq  l' $

 $ G_{k+1} (\tuple{l, F_1, \ldots, F_n}, (\act_1, \ldots, \act_n), l') \defeq $

~~~~$ \tuple{l', H_{k, 1} (F_1, \act_1, \obsi{1}{l'}), \ldots, H_{k, n} (F_n, \act_n, \obsi{n}{l'})} $

$ H_{k, \agi} (F_i, \act_\agi, o_\agi) \defeq  $

~~~~$ \setdef{G_k (t^{(k)}, \act, l)}
                      {t^{(k)} \in F_\agi, \act (\agi) = \act_\agi, (\mathit{root} (t^{(k)}), \act, l) \in \Delta, l \in o_\agi} $
\end{definition}

The above definition is slightly more general than the original one
from~\cite{DBLP:journals/iandc/Meyden98}, since the game model 
presented there does not involve named actions. 


\end{document}